\documentclass[twocolumn]{aastex61}

\usepackage{graphicx}
\usepackage{txfonts}
\usepackage{natbib}

\received{04/12/2017}
\accepted{12/07/2018}
\submitjournal{ApJ}
\shorttitle{Evolution of coronal loop structure}
\shortauthors{Goddard et al.}

\begin{document}
\title{Evolution of the transverse density structure of oscillating coronal loops inferred by forward modelling of EUV intensity}

\author[0000-0003-0240-0465]{C. R. Goddard}
\email{c.r.goddard@warwick.ac.uk}
\affiliation{Centre for Fusion, Space and Astrophysics, Department of Physics, University of Warwick, CV4 7AL, UK}

\author[0000-0003-1529-4681]{P. Antolin}
\affiliation{School of Mathematics and Statistics,
University of St Andrews, St Andrews, Fife,
KY16 9SS, UK}

\author[0000-0002-0338-3962]{D. J. Pascoe}
\affiliation{Centre for Fusion, Space and Astrophysics, Department of Physics, University of Warwick, CV4 7AL, UK}
\affiliation{Centre for Mathematical Plasma Astrophysics, Mathematics Department, KU Leuven, Celestijnenlaan 200B bus 2400, B-3001 Leuven, Belgium}

\begin{abstract}
Recent developments in the observation and modelling of kink oscillations of coronal loops have led to heightened interest over the last few years. The modification of the Transverse Density Profile (TDP) of oscillating coronal loops by non-linear effects, in particular the Kelvin-Helmholtz Instability (KHI), is investigated. How this evolution may be detected is established, in particular, when the KHI vortices may not be observed directly. A model for the loop's TDP is used which includes a finite inhomogeneous layer and homogeneous core, with a linear transition between them. The evolution of the loop's transverse intensity profile from numerical simulations of kink oscillations is analysed. Bayesian inference and forward modelling techniques are applied to infer the evolution of the TDP from the intensity profiles, in a manner which may be applied to observations. The strongest observational evidence for the development of the KHI is found to be a widening of the loop's inhomogeneous layer, which may be inferred for sufficiently well resolved loops, i.e $>$ 15 data points across the loop. The main signatures when observing the core of the loop (for this specific loop model) during the oscillation are: a widening inhomogeneous layer, decreasing intensity, an unchanged radius, and visible fine transverse structuring when the resolution is sufficient. The appearance of these signatures are delayed for loops with wider inhomogeneous layers, and quicker for loops oscillating at higher amplitudes. These cases should also result in stronger observational signatures, with visible transverse structuring appearing for wide loops observed at SDO/AIA resolution.
\end{abstract}


\section{Introduction} \label{sec:intro}

Kink (or transverse) oscillations of coronal loops have been intensively studied over the last two decades since their detection with the Transition Region And Coronal Explorer (TRACE) \citep{1999SoPh..187..229H} in 1999 \citep{1999ApJ...520..880A, 1999Sci...285..862N}. Many examples of standing kink modes have been clearly observed \citep[e.g][]{2012A&A...537A..49W, 2016A&A...585A.137G,2016SoPh..291.3269S, 2017ApJ...842...99L} with the enhanced spatial and temporal resolution of the Atmospheric Imaging Assembly (AIA) on board the Solar Dynamics Observatory (SDO) \citep{2012SoPh..275...17L}. The accepted mechanism for the rapid damping of these oscillations is resonant absorption \citep[e.g.~recent review by][]{0741-3335-58-1-014001}.

Kink oscillations can be used to perform seismology and obtain estimates for the local plasma parameters \cite[e.g][]{2001A&A...372L..53N,2012RSPTA.370.3193D,2013A&A...552A.138V}, which can aid studies of other processes in the Suns atmosphere. Recently in \cite{2016A&A...589A.136P,2017A&A...600A..78P} this approach was updated to include the proposed Gaussian and exponential damping regimes \citep{2013A&A...551A..39H,2013A&A...551A..40P}, which makes the inversion problem well posed when the switch between the two regimes can be observed. This is based on a simplified model of the Transverse Density Profile (TDP) of the loop, described by a uniform core with an inhomogeneous layer where the density varies linearly between the background and internal density. This analytic description is also subject to the thin boundary layer approximation, which has been shown to vary the damping rates due to resonant absorption by a factor of 2 for thick non-uniform layers \citep{2004ApJ...606.1223V}.

An agreement between the seismologically determined TDP and the TDP inferred from the transverse intensity profile was found in \cite{2017A&A...600L...7P}. This was extended in \cite{2017A&A...605A..65G} to perform a statistical study of the TDPs of 233 coronal loops. In many cases there was evidence for a Gaussian transverse density profile, or thick linear inhomogeneous layer. The Bayesian inference approach allows the different density profiles to be quantitatively compared, and also allows robust estimation of the uncertainties of the model parameters. This study indicated that loops may have thicker boundary layers than is typically assumed, or even constantly varying TDPs, subject to the limitations and simplifications discussed. Additionally in \cite{2018ApJ...860...31P} this technique was applied to an oscillating coronal loop to infer the time evolution of the density profile model parameters.

Seismological studies assume that the TDP of the loop remains constant during the oscillations, making it important to understand any changes which do occur. There are many effects and non-linear mechanisms which can cause the TDP to vary, which will in turn modify the observed damping behaviour. Large amplitude kink waves have been shown to produce plasma flows along the field, and the ponderomotive force can cause accumulation of density at the loop top \cite[e.g][]{2004ApJ...610..523T, 2009PhPl...16g2115C, 2012A&A...544A.127V}. The effect of a time-varying cross-section was recently investigated analytically in \cite{2017A&A...602A..50R}. In \cite{2016A&A...590L...5G} it was observed that the quality factor of kink oscillations decreases as the oscillation amplitude increases, indicating that finite amplitude effects are playing some role in modifying the damping time and/or period. This could be due to effects which modify the structure of the loop at high amplitudes. A qualitatively similar dependence was found in \cite{2016A&A...595A..81M}, where non-linear effects such as the growth of the KHI instability were found to modify the damping profile of the kink mode at high amplitudes.

The Kelvin-Helmholtz instability (KHI) \citep{1984A&A...131..283B, 1994GeoRL..21.2259O} has been shown to occur in numerical simulations within the inhomogeneous layer of oscillating loops or prominences due to the shear flows, redistributing both the density and temperature of the plasma \citep[e.g][]{2008ApJ...687L.115T, 2010ApJ...712..875S, 2014ApJ...787L..22A}. In particular \cite{2016ApJ...830L..22A, 2017ApJ...836..219A} included forward modelled EUV emission from a loop subject to the KHI instability, noting how the oscillation can appear decayless under certain circumstances, and how seismology can be performed based on the phase mixing which takes place. The loop's structure can also evolve during oscillations if they contain unresolved sub-structure or multi--threadedness. Recently \cite{2016ApJ...823...82M} showed that transverse oscillations in loops with substructure cause the strands to merge and produce a more homogeneous density structure. The KHI vortices generated in these simulations are often referred to as Transverse Wave Induced KHI (TWIKH) rolls.

In this paper the technique used in \cite{2017A&A...600L...7P,2018ApJ...860...31P} and \cite{2017A&A...605A..65G} is applied to forward modelled EUV emission from numerical simulations of kink oscillations \citep{2017ApJ...836..219A}. In contrast to \cite{2017ApJ...836..219A}, the effect of the development of KHI on the parameters of a TDP model inferred from the EUV emission is investigated. The variation of the parameters may be detected even when the resolution is not sufficient to resolve the complex substructure generated by the TWIKH rolls. In addition, simulations with a larger amplitude of oscillation, and a loop with a larger inhomogeneous layer are analysed and compared. In Sect.~\ref{sec:numdat} the numerical models and data are described. In Sect.~\ref{sec:intfit} the analysis is described and the effects of resolution and noise are explored. The obtained temporal evolution is presented in Sects.~\ref{sec:m1} and \ref{sec:m2m3}, and further discussion and summary are given in Sect.~\ref{sec:disc}.


\begin{figure}
\centering
\includegraphics[width=9.5cm]{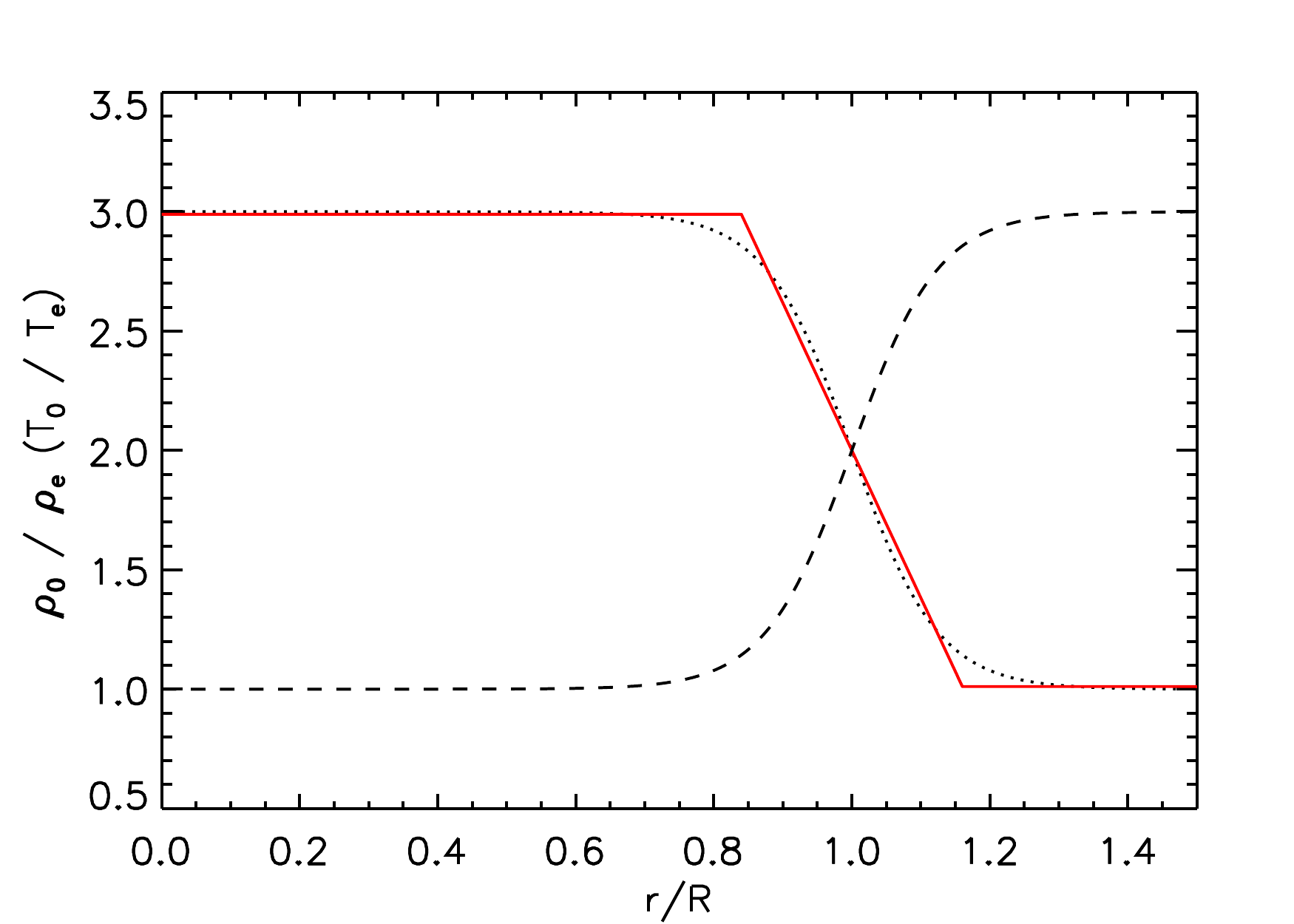} 
\caption{The initial density and temperature profile of the loop in models M1 and M3 (dotted and dashed respectively) and an approximation of the density profile using Model $L$ (solid red). The $x$ axis is given as the radial coordinate $r$ divided by the loop minor radius $R$, defined to occur halfway though the loop's inhomogeneous layer. The fit shown by the red line gives a value for the width of the inhomogeneous layer of $\epsilon$ = 0.32.}
\label{dprof}
\end{figure}

\section{Numerical data}\label{sec:numdat}

\subsection{Loop model}
The numerical data analysed is based on the modelling by \cite{2016ApJ...830L..22A,2017ApJ...836..219A}. One of the same simulations is used, presented as Model 1 in \cite{2017ApJ...836..219A}, which will be called M1. The parameters of this model are given in Table~\ref{tab:var}. 
The numerical model used is a straight loop in MHD equilibrium. The magnetic field is uniform throughout the domain at 22.8 G and the loop has a temperature of 1 MK, cooler that the background plasma by a factor of 3. To maintain pressure balance the density in the loop is 2.5x10$^{-15}$ g cm$^{-3}$, 3 times the external density. A smooth transition layer connects the internal plasma with the external plasma (see Figure~\ref{dprof}). At time $t =$ 0 a velocity perturbation mimicking the fundamental kink mode, with longitudinal wavenumber $kR= \pi R/L \approx 0.015$ (where $R$ is the loop's minor radius and $L$ is the length of the loop) is applied to the loop along the $x-$direction. This has a velocity amplitude of $\approx15~$km~s$^{-1}$ in the case of M1 (leading to a displacement of $A_{0}\approx0.4~R$).

The numerical scheme is the CIP-MOCCT code \citep{1999ApJ...514..493K}, and solves the MHD equations of mass conservation, momentum, magnetic field induction and energy (for an ideal fully ionised plasma), excluding gravity, radiative cooling and thermal conduction. The simulation is close to ideal with no explicit resistivity or viscosity. The numerical box is $512 \times 256 \times 50$ grid points in the $x, y$, and $z$ directions, respectively. Due to the symmetric properties of the kink mode, only half of the plane in $y$ and half of the loop are modelled (from $z = 0$ to $z = 100~R$). Symmetric boundary conditions are set in $y$ and $z$, for $x$ periodic boundary conditions are imposed. In order to minimise the influence from side boundary conditions (along $x$ and $y$), the spatial grids in $x$ and $y$ are non-uniform with exponentially increasing values for distances beyond the maximum displacement. The maximum distance from the centre in $x$ and $y$ is $\approx16~R$. The spatial resolution at the loop's location is $0.0156~R = 15.6~$km. From a parameter study, the effective Reynolds and Lundquist numbers in the code are estimated to be of the order of $10^4$--$10^5$ \citep{2015ApJ...809...72A}. The temporal variation in temperature in these models is therefore mostly due to adiabatic effects. For further numerical detail see \cite{2017ApJ...836..219A} and the references therein.

The loop model used corresponds to a cool and dense loop, in hotter rarefied surroundings. It is unknown how well this model applies to the dense, largely isothermal loops typically detected with EUV imagers at 171 $\AA$ due to the difficulty in defining the temperature of the ambient background, and the variation between active regions. Also, the magnetic field strength of neighbouring field lines is taken to be constant, which is not necessarily the case and would change the requirements for pressure balance across the loop. Since the same forward modelling and Bayesian inference technique used on observations in previous studies is used here, the linear transition layer model is applied to approximate the actual density profile, to define it's parameters. This model ($L$) is defined as;

\begin{eqnarray}
\rho \left( r \right) = 
\begin{array}{cr}
A,  & \vert r \vert \le r_1 \\
A \left( 1 - \frac{r-r_1}{r_2-r_1} \right), & r_1 <  \vert r \vert \le r_2 \\
0, & \vert r \vert > r_2 \\
\end{array}.
\label{eq:ml}
\end{eqnarray}
where $r_1=R_{L} \left( 1 - \epsilon/2 \right)$, $r_2=R_{L} \left( 1 + \epsilon/2 \right)$, and $\epsilon = l/R_{L}$ is the transition layer width $l$ normalised to the minor radius $R_{L}$ and defined to be in the range $\epsilon \in \left[ 0,2 \right]$.
Finally, $A$ is the density enhancement, $A = \rho_0 - \rho_e$. In our analysis $A$ will be normalised since it does not reflect the absolute value of the density contrast discussed further in Section~\ref{sec:intfit}. The fitting of the original density profile with model $L$ is shown in Figure \ref{dprof}, this gives the values of $\epsilon$ given in the Table~\ref{tab:var}.  

\begin{table}[]
\centering
\caption{Parameter values for M1, M2 and M3.}
\label{tab:var}
\begin{tabular}{|l|lll|}
\hline
Parameter                                     & \multicolumn{1}{l|}{M1} & \multicolumn{1}{l|}{M2} & \multicolumn{1}{l|}{M3} \\ \hline
$\rho_{0} (0.5 \times 10^{9}m_{p} (gcm^{-3})$ &                         & 3                       &                         \\ \hline
$\rho_{0}/\rho_{e}$                           &                         & 3                       &                         \\ \hline
$T_{0} (MK)$                                  &                         & 1                       &                         \\ \hline
$T_{0}/T_{e}$                                 &                         & 1/3                     &                         \\ \hline
$B_{0}/B_{e}$                                 &                         & 1                       &                         \\ \hline
$\epsilon$                                    & 0.32                    & 0.64                    & 0.32                    \\ \hline
$A_{0}/R$                                     & 0.4                     & 0.4                     & 0.8                     \\ \hline
$R$ (pixels)                                  &                         & 64                      &                         \\ \hline
$R_{1}$ (pixels)                              &                         & 32                      &                         \\ \hline
$R_{2}$ (pixels)                              &                         & 8                       &                         \\ \hline
\end{tabular}
\end{table}

Two additional models are considered, M2 and M3. In M2 the effect of changing the width of the inhomogeneous layer is investigated, the parameters are consistent with those given for M1, but with twice the layer width ($\epsilon$). In M3 the effect of changing the amplitude of the oscillation is investigated, the parameters are consistent with those given for M1, but with double the value of the initial amplitude ($A_0$). 

\subsection{TD maps}
\label{init_td}

The numerical data is provided as a Time--Distance (TD) map which has been forward modelled to EUV emission from the 171 $\AA$ AIA channel in the same manner as described in \cite{2017ApJ...836..219A}. The original size of the numerical data is $512$ pixels in the direction across the loop, and in the vicinity of the loop one pixel corresponds to 15.6~km. In the plotted TD maps one time--step on the temporal axis corresponds to 6.96 s. However, the model can be considered as scale free for the analysis presented here, and so spatial co-ordinates will be given in terms of pixels, and temporal coordinates will be given normalised to the oscillatory period of the loop, 255 s.

This spatial domain is reduced to 256 (M1), 256 (M2), and 296 (M3) to avoid unnecessary pixels around the loop being included, increasing the run time of the Bayesian inference. The data is then interpolated to lower spatial resolutions for analysis, corresponding to 128 (R1) and 32 (R2) pixels in the spatial plane of the TD map. The actual loop minor radius (determined from the density profile) $R$ is 32 px at R1 and 8 px at R2 when the TDP is fit directly with model $L$. R2 corresponds to a loop with an apparent minor radius (when estimated from the intensity profile) of $\approx$ 5 Mm if it was observed with AIA (where one pixel is 0.6 arcsec, and the effect of LOS integration and the PSF is included). This is similar to the example analysed in \cite{2017A&A...600L...7P}, which had an apparent radius of 4.5 Mm. R1 corresponds to an unrealistically wide loop at AIA resolution (the observed radius would be $\approx$ 20 Mm), however it allows us to explore the effect that the higher spatial resolution has on the results, and also what may be observed with higher resolution instruments in the future (i.e increased pixel count across the loop). The TD maps are left unchanged in the temporal coordinate.

Noise is added to the TD maps to simulate noise in the EUV intensity data from imaging instruments. The noise, which is a function of the intensity, is generated according to the equation

\begin{equation}
td_N[i] =  td_0[i] + N \times max(td_0)\times(1 + td_0[i]/max(td_0)) + BG
\label{}
\end{equation}

where $N$ is the noise, $td_N$ is the TD map with the added noise, $td_0$ is the original TD map intensity and $[i]$ denotes a particular pixel. A level of background intensity is also added ($BG$). $N$ is given by $N$ = $A_N$ $\times$ $Rnd$, where $A_N$ is used to prescribe the amplitude of the noise and $Rnd$ is a randomly generated number from a Poisson distribution. 

Finally the effect of the instrumental Point Spread Function (PSF) of the 171 $\AA$ AIA channel \citep{aiapsf} is approximated by smoothing the TD map in the spatial coordinate with a Gaussian. The Gaussian width used is $\sigma$ = 1.019 pix. This is kept constant at the different spatial resolutions used for simplicity. Other effects which may vary the resolving power of an EUV imaging instrument instrument are neglected.

\begin{figure}
\centering
\includegraphics[width=8cm]{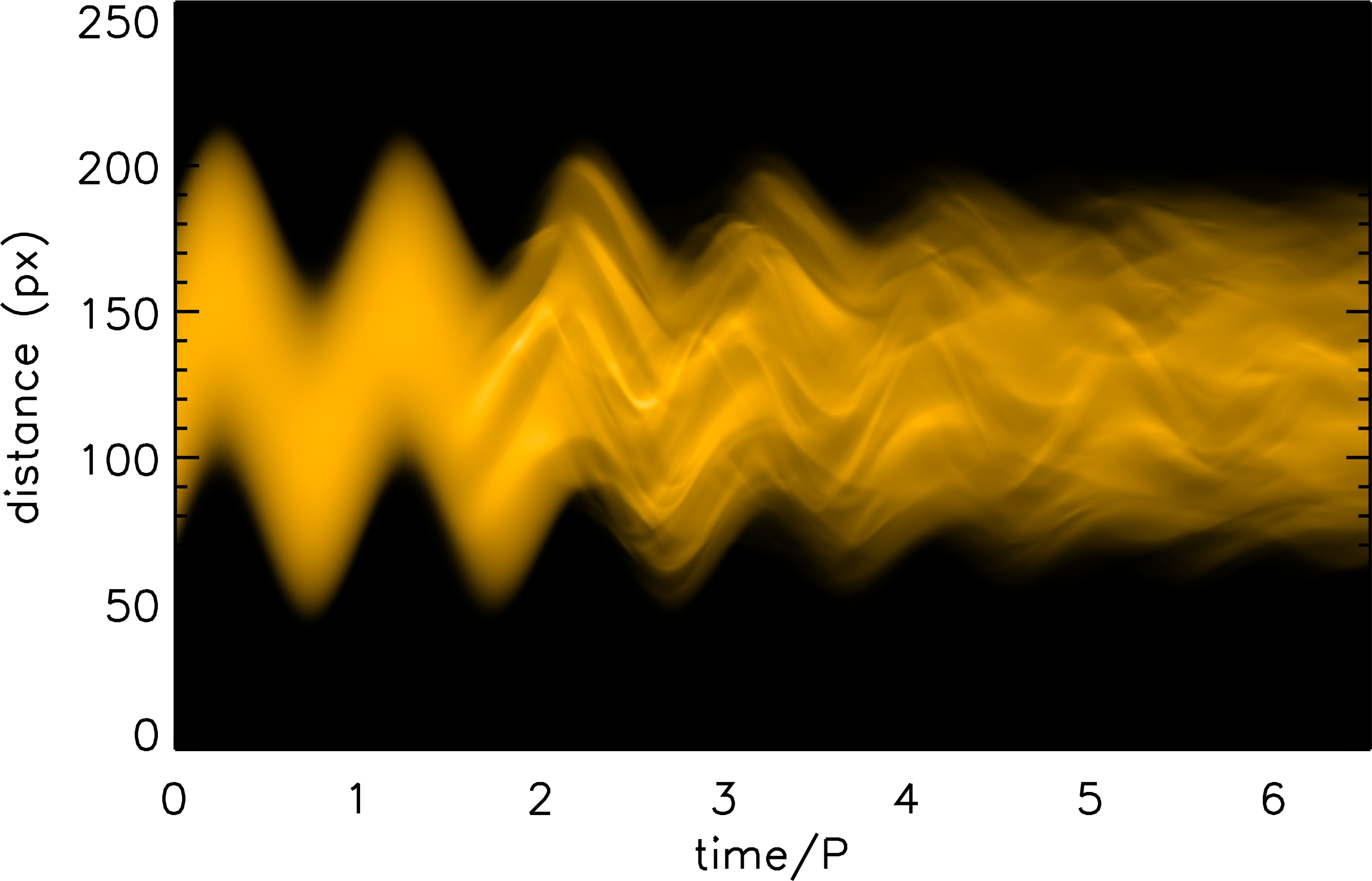}
\includegraphics[width=8cm]{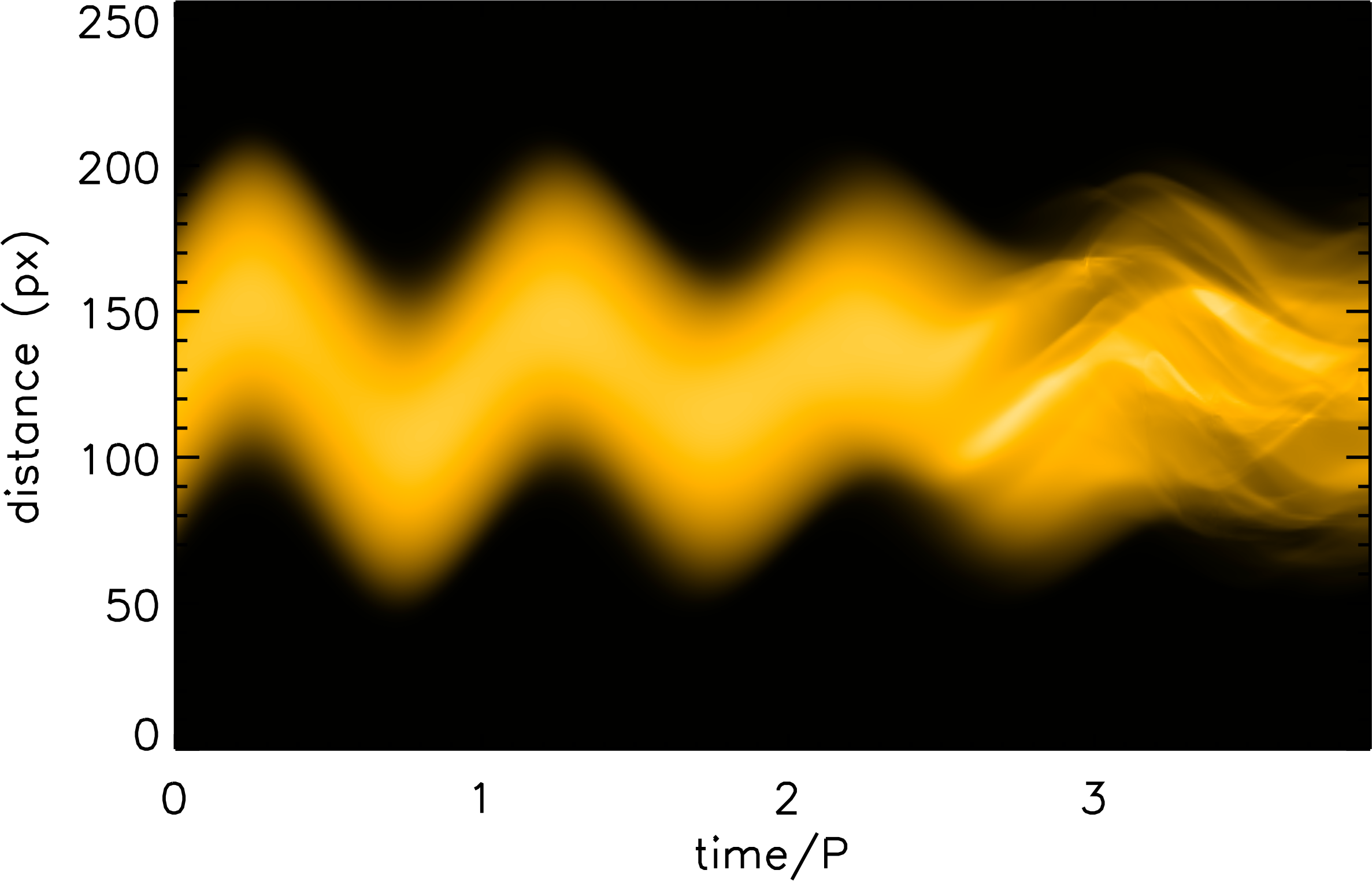}  
\includegraphics[width=8cm]{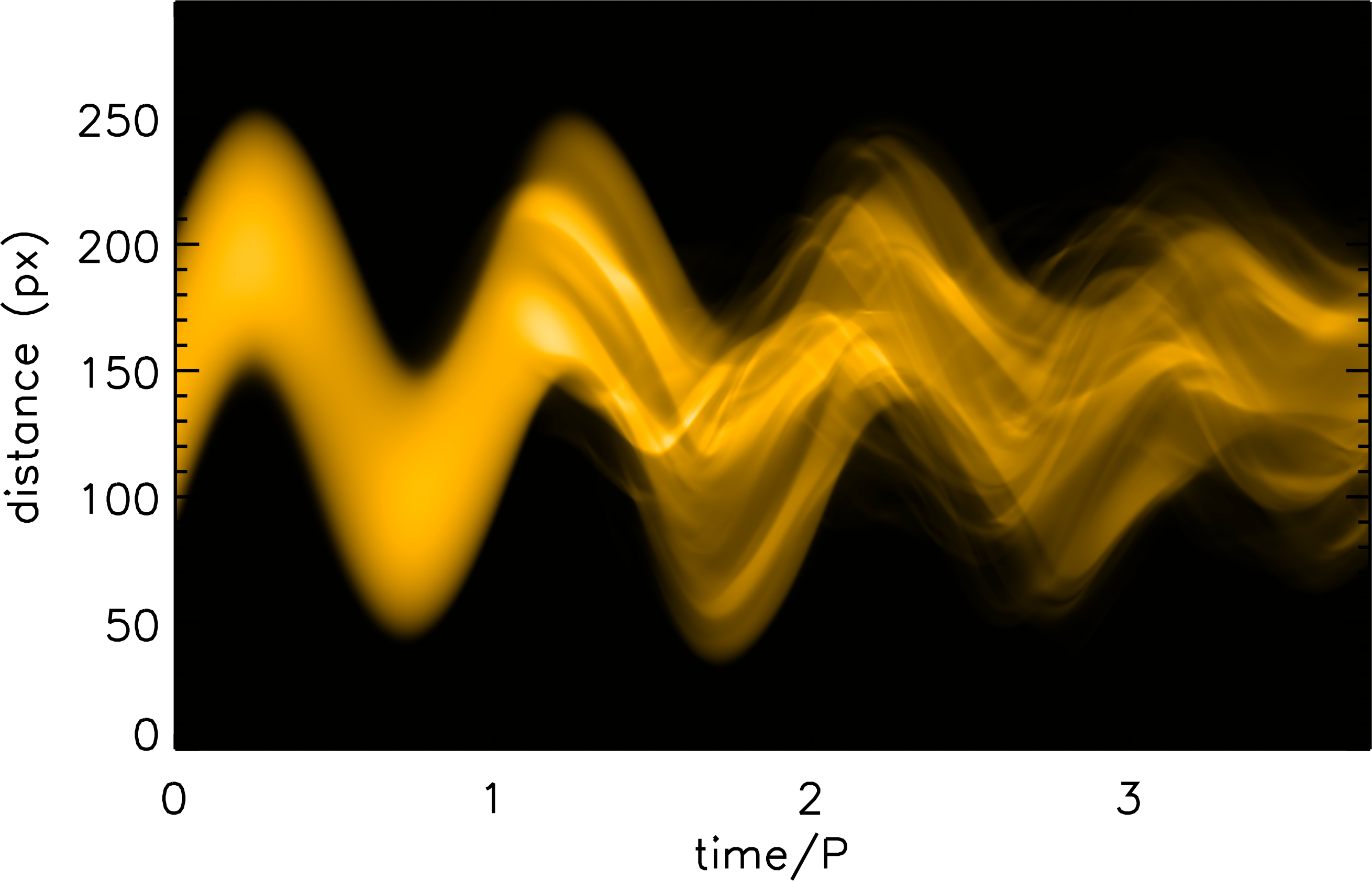} 
\caption{Low amplitude (M1), large inhomogeneous layer (M2) and high amplitude (M3) TD maps (top to bottom) at $171${\AA} forward modelled from numerical simulations of oscillating coronal loops. The horizontal axis is time normalised to the period of the kink oscillation, $P$. The vertical axis corresponds to distance across the loop in numerical pixels.}
\label{orig_tdmaps}
\end{figure}

In Fig. \ref{orig_tdmaps} the TD maps corresponding to M1, M2 and M3 are plotted at the original spatial resolution of the simulation. In the top panel the disruption and fine structuring of the initially homogeneous intensity profile is seen to occur as the KHI instability develops from around $t \approx 2 P$, where $P$ is the period of the kink oscillation. The second panel is M2, where the onset of the KHI is delayed by the larger inhomogeneous layer. This occurs, despite the enhanced efficiency of the resonant absorption, as the less steep initial density gradient increases the phase mixing time scale meaning it takes longer for sharp gradients in density and velocity to develop \citep[e.g][]{1991SoPh..134..111U,2017A&A...601A.107P, 2017arXiv171206955T}. The growth time of the KHI is therefore reduced until later in the simulation, $t \approx 3 P$. M3 is shown in the lower panel. The KHI instability develops at $t \approx 1 P$, and by the final frame the loop is the most disrupted out of the three simulations, due to the higher oscillation amplitude generating the KHI more efficiently. An exploration of the effect of oscillation amplitude on the development of the KHI instability was made in \cite{2016A&A...595A..81M}.

It should be noted that M2 and M3 have approximately half the oscillation cycles of M1 as the generated spatial scales are on the order of the grid size beyond this time. The large energies of the vortices at these scales would require additional treatment to ensure the code remains numerically stable, meaning that the results would not be directly comparable between the different models, and is beyond the scope of this work.

\section{Intensity profile fitting}\label{sec:intfit}

\begin{figure*}
\centering
\includegraphics[width=5.9cm]{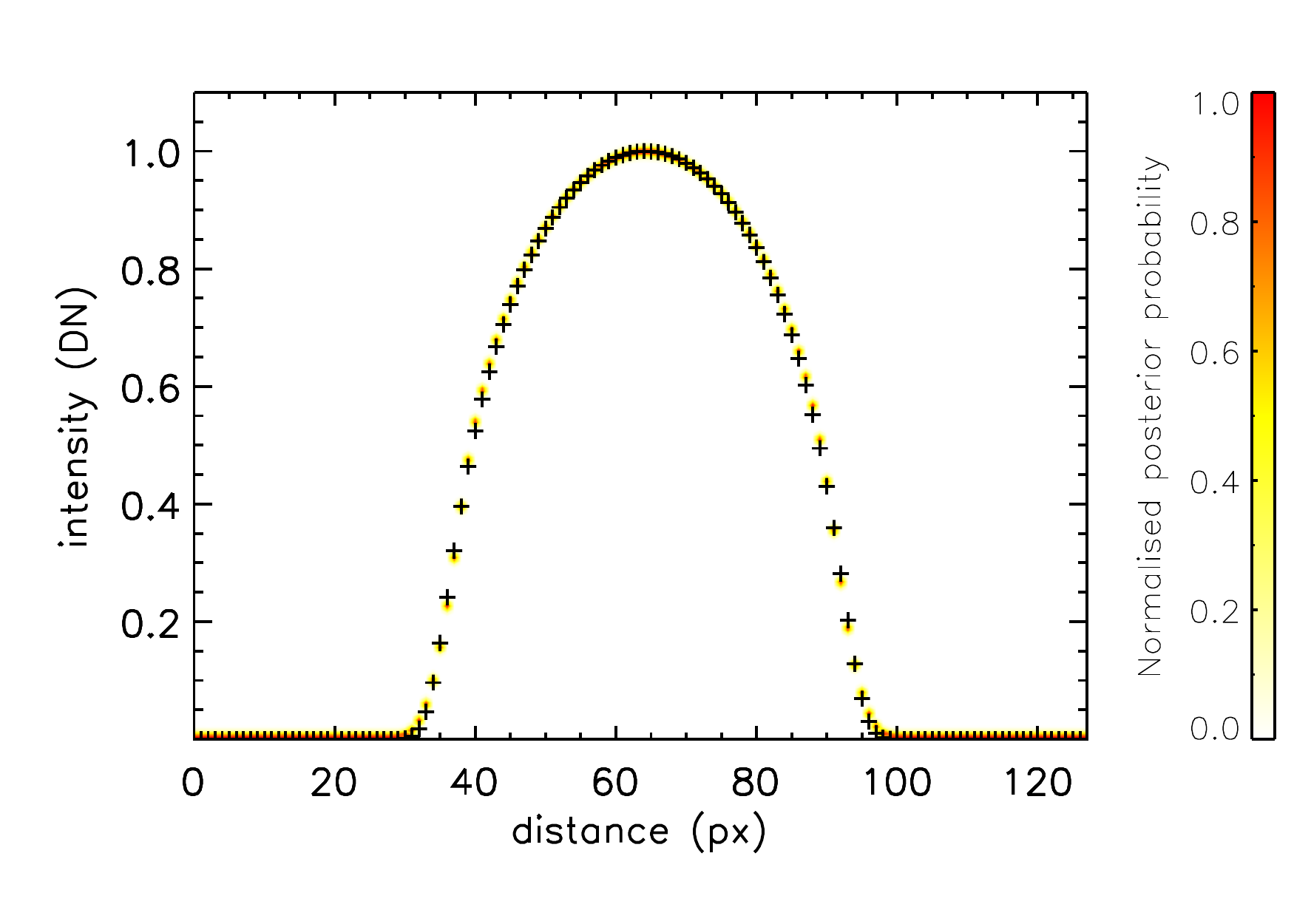}  
\includegraphics[width=5.9cm]{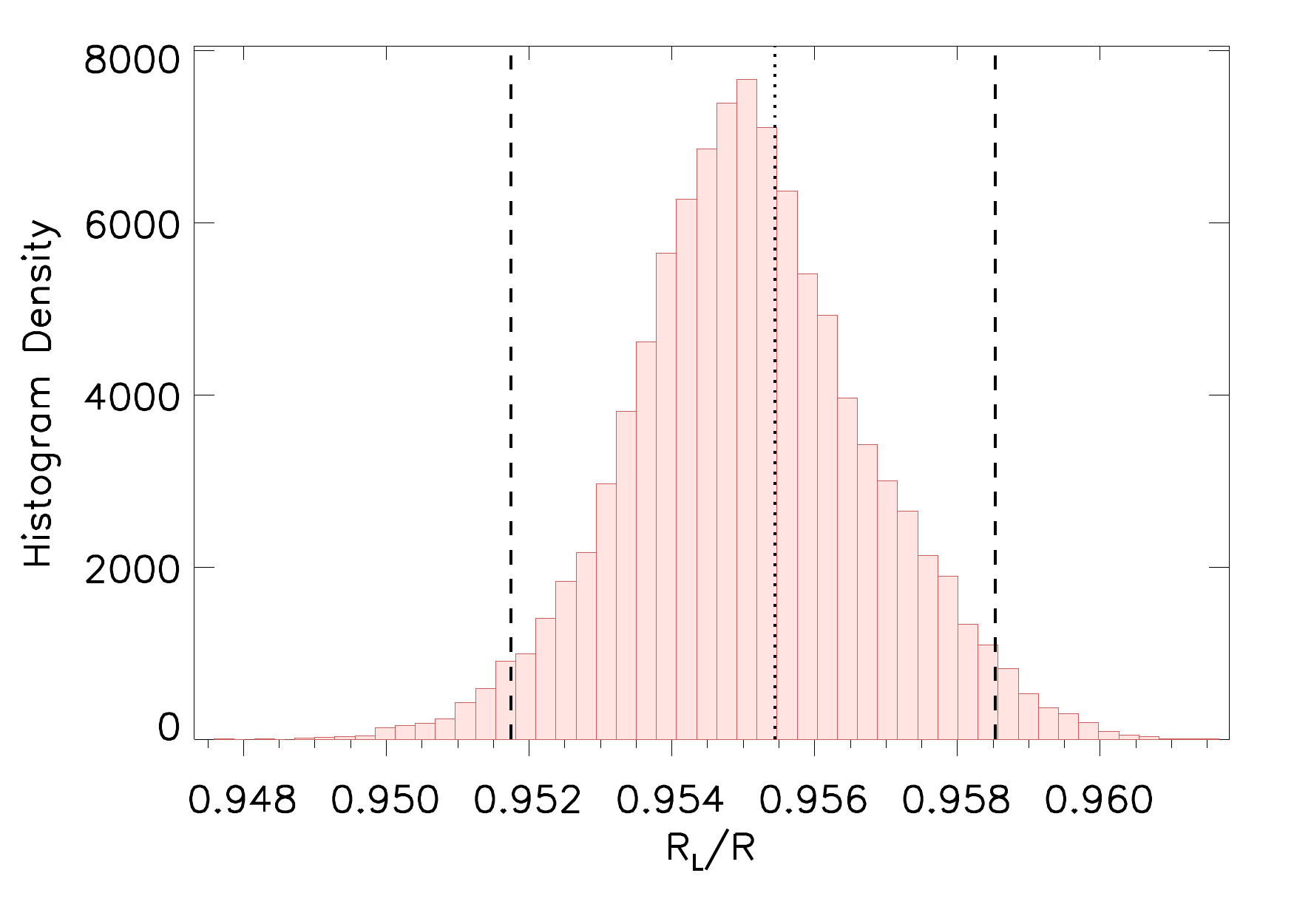}
\includegraphics[width=5.9cm]{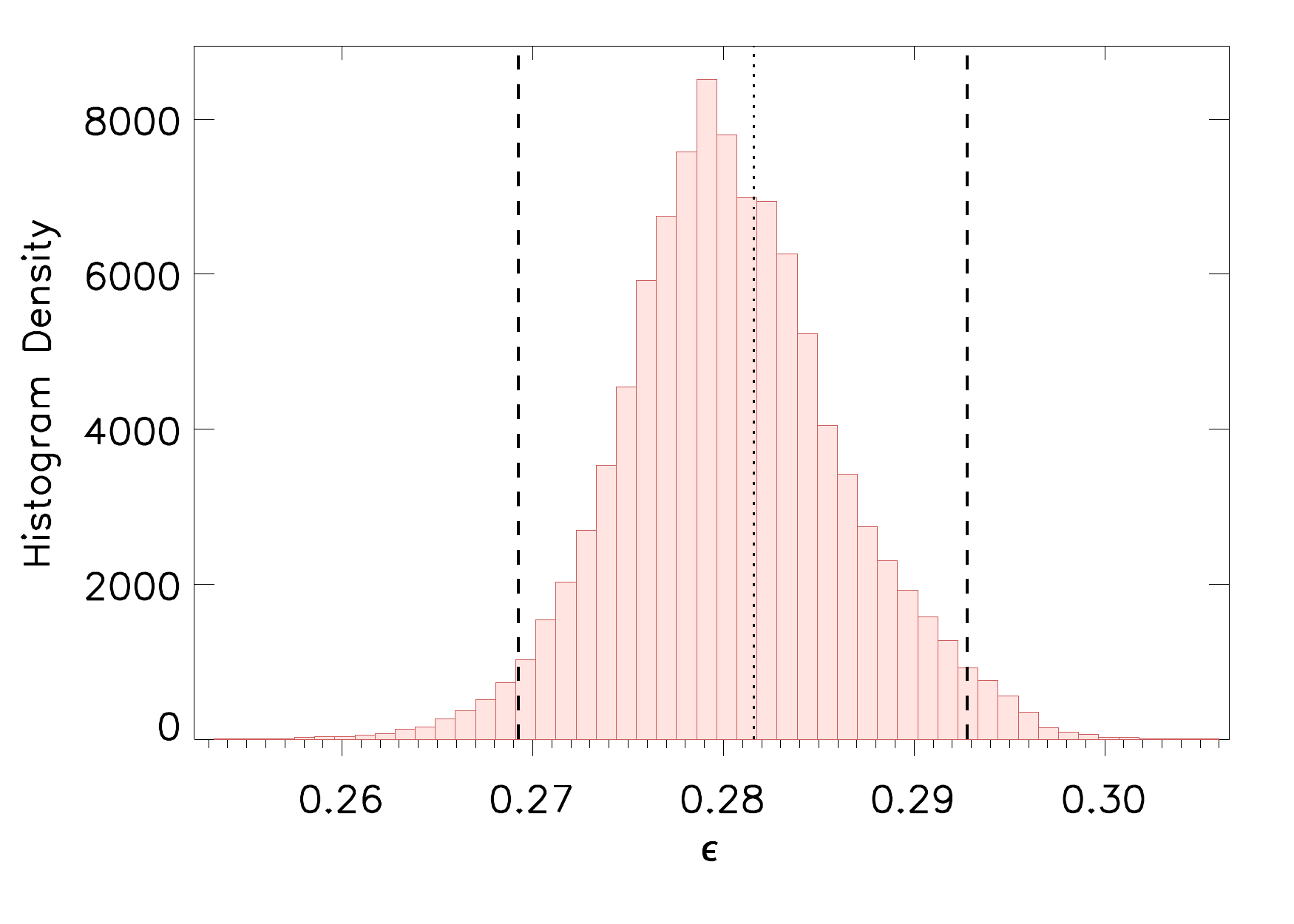}
\caption{The initial intensity profile for M1 (top left) at spatial resolution R1, with the normalised posterior probability for each point plotted in the background. The other panels show the probability density for two parameters of density profile model L, the radius ($R_{L}/R$) and layer width ($\epsilon$), corresponding the intensity profile on the left. The radius has been normalised by the actual radius used in the simulation. The two dashed lines correspond to the upper and lower 95 \% confidence intervals and the dotted line corresponds to the Maximum A posteriori Probability (MAP) values of the parameters.}
\label{m1_profits}
\end{figure*}

\begin{figure*}
\centering
\includegraphics[width=8.9cm]{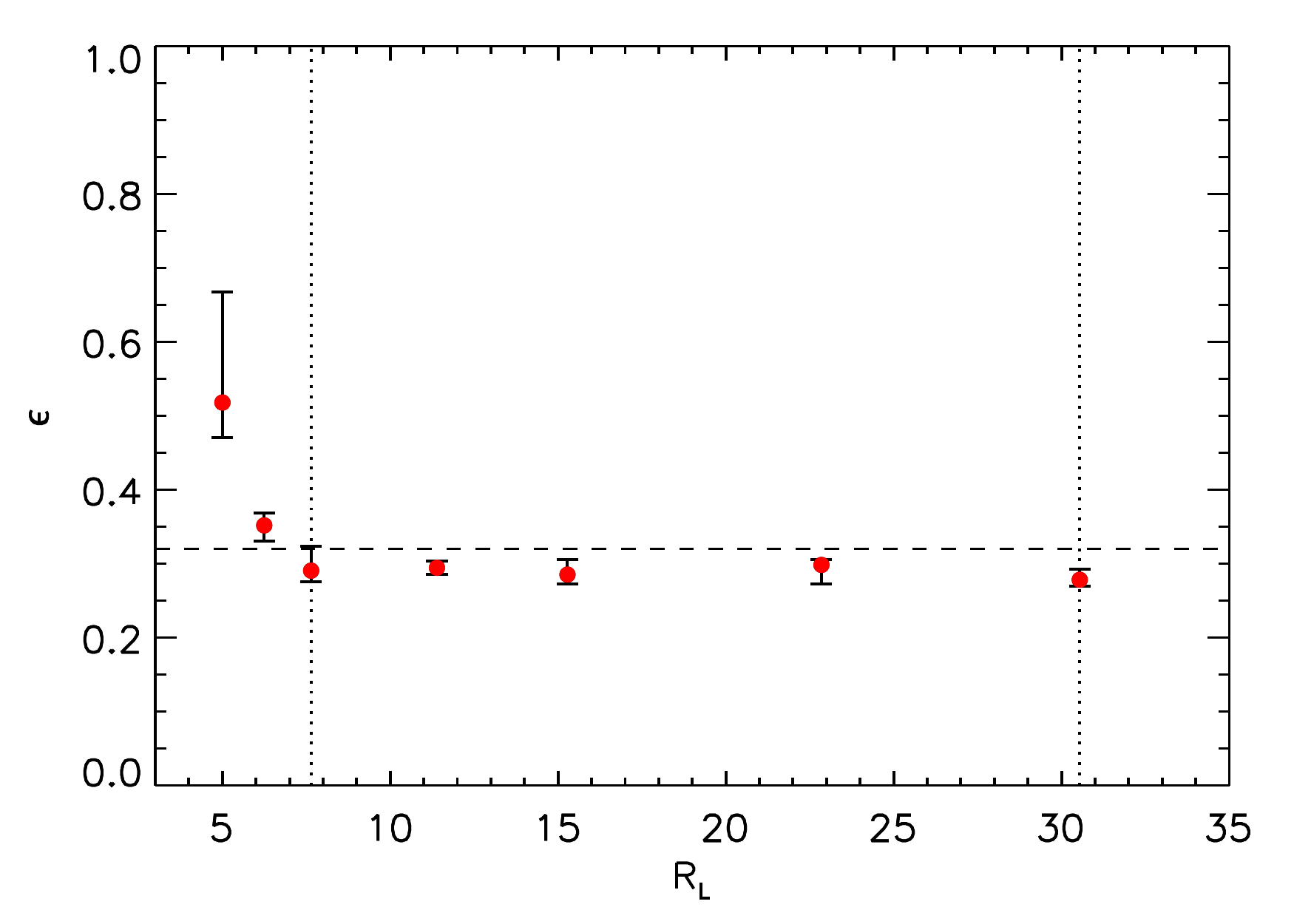} 
\caption{MAP value of $\epsilon$ and it's uncertainty plotted as a function of the radius $R_L$ obtained from Model $L$ and the transverse intensity profile for M1. The vertical dotted lines correspond to R1 and R2, the resolutions used in the time series fitting.}
\label{res_test}
\end{figure*}

Recently there have been several studies related to kink oscillations which have used the Bayesian methodology \citep[e.g][]{2011ApJ...740...44A,2015ApJ...811..104A,2017ApJ...846...89M}.
\citet{2017arXiv170908372A} also produced a review of coronal seismology using Bayesian analysis.
The same Bayesian inference and MCMC sampling approach is used, as previously applied to obtain the TDPs of coronal loops in \cite{2017A&A...600L...7P,2017A&A...605A..65G} and the evolution over time in \cite{2018ApJ...860...31P}. This approach allows robust estimation of the uncertainties on the inferred parameters, clearly identifies redundant parameters and allows quantitative model comparison to be performed. This procedure is well suited to analysing the numerical data presented here to determine what features of the density profile evolution can be detected for oscillating loops observed with SDO/AIA and future EUV imagers which better resolved the transverse structure of coronal loops.

Model $L$ (the linear transition layer model) for the density profile will be used, as given in Equation 1. \cite{2018ApJ...860...31P} considered seven different transverse density profiles which have been used to describe coronal loops and demonstrated that the linear transition layer profile described the widest range of possible structures, being able to approximate homogeneous, partially inhomogeneous and completely inhomogeneous density profiles. The loop in the numerical modelling was not isothermal, and as such the cooler core is well seen at 171 $\AA$ channel, and the hot boundary is better seen in a hotter channel. The transverse intensity profile therefore becomes double peaked in the hotter channel \citep[see][]{2017ApJ...836..219A}, and is not suitable for this analysis (where an isothermal approximation is made). However it was noted in \cite{2017ApJ...836..219A} that the structure in the hotter channel may appear homogeneous at very low resolutions, but at such low resolutions there are not enough data points across the loop to infer meaningful information about the transverse structure. Often non-flaring loops observed at 171 $\AA$ appear to be relatively isothermal, and as such are not seen well in other channels \citep[e.g][]{2011ApJ...732...81A, 2017A&A...605A..65G}. The analysis is restricted to the simulated emission at 171 $\AA$.

The intensity profile is forward modelled from the density profile as described in the previous studies, using a cylindrical cross-section, an approximation of the PSF of the given AIA channel ($\AA$) and an isothermal approximation. The same Bayesian inference procedure is then performed to obtain Maximum A posteriori Probability (MAP) values and their uncertainties (the 95 $\%$ credible interval) for the parameters of Model L from the numerical TD map. In principle this could be performed with least-squares fitting, however the estimation of the parameter uncertainties would be less robust and it would give no information about parameter redundancy.

It is important to note that the values of $A$ that are obtained do not relate to the actual density enhancement, since the intensity profiles are normalised, and an arbitrary level of background intensity is added before the analysis that follows. The intensity contrast is also affected by the LOS depth over which the EUV emission is integrated. When analysing numerical data for which the size of the numerical domain is known it is possible to recover the actual density contrast. However this is not generally applicable to observations for which the integration depth is unknown.

In the top left of Figure \ref{m1_profits} the transverse intensity profile of the loop for M1, at resolution R1, at $t$ = 0 is shown. No noise has been added, but the approximation of the instrumental PSF has been applied. The background contour plot corresponds to the  normalised predictive posterior probability density for each data point, i.e how likely it is that the data point will lie in a certain position given the corresponding model parameters. The other two panels show the posterior probability density for  $R_{L}/R$ (the inferred radius normalised by the actual radius) and $\epsilon$ from model $L$, also at $t$ = 0. It can be seen that despite working backwards from the intensity profile with the PSF applied, and the isothermal approximation, the obtained value of $\epsilon$ is sufficiently close to the values obtained from direct fitting of the density profile (0.32). The dashed lines correspond to the credible intervals for the MAP value for that parameter (dotted line). The obtained value of $R_{L}/R$ is 5\% less than 1, showing that the radius is slightly underestimated because only the plasma which emits in the 171 $\AA$ channel is analysed.

\subsection{Resolution test}

\begin{figure*}
\centering
\includegraphics[width=8.9cm]{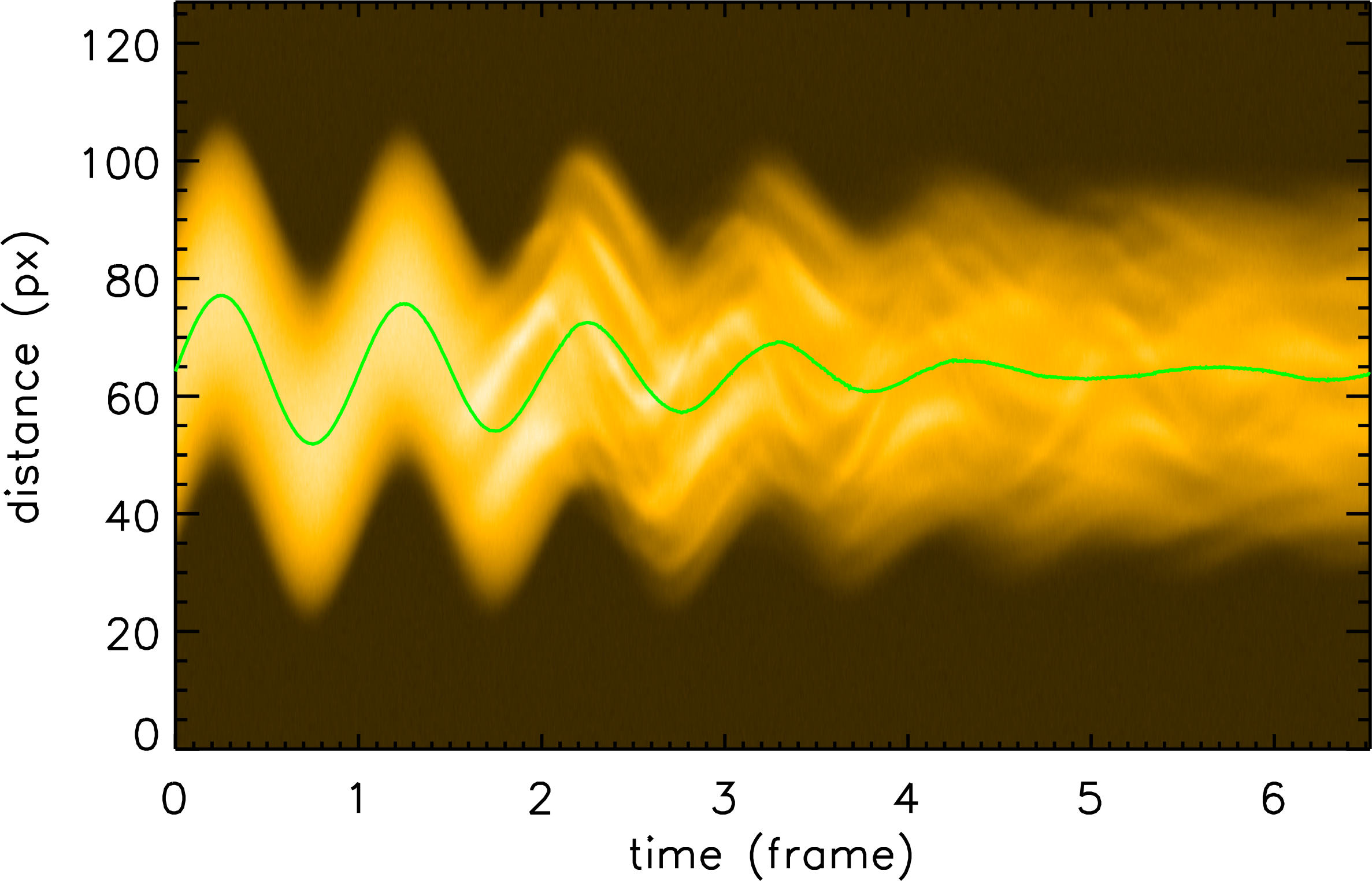}
\includegraphics[width=8.9cm]{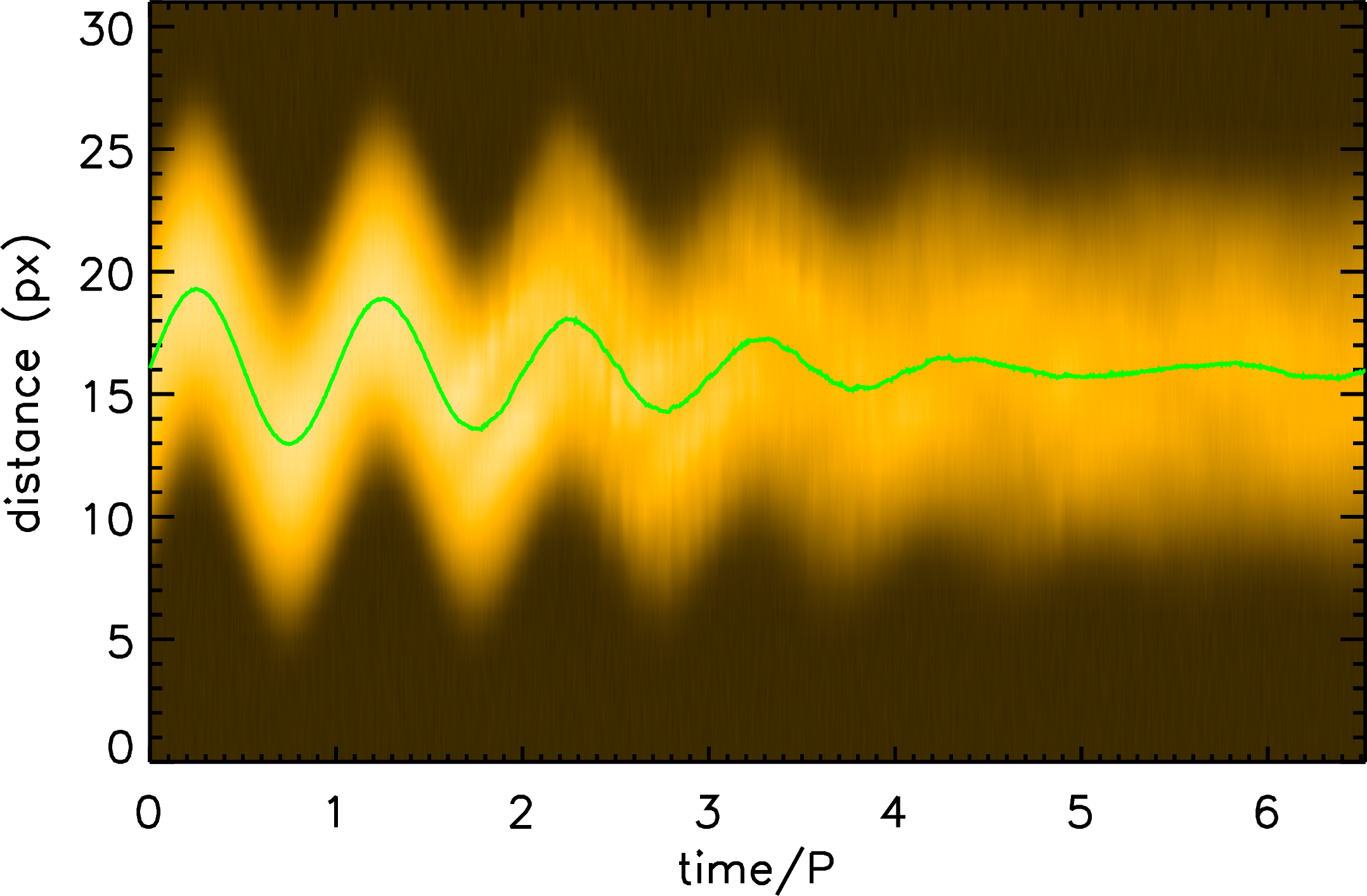}
\caption{TD maps for M1 at resolutions R1 and R2. The overplotted green lines correspond to the MAP value of the loop centre position at each time. The left column corresponds to resolution R1 ($R$ = 32 pixels) and the right column to a four times lower resolution, R2 ($R$ = 8 pixels).}
\label{m1_tdmaps}
\end{figure*}

The effect of lowering the resolution of the numerical data before analysis on the inferred value of $\epsilon$ is tested. Since the 171 $\AA$ AIA PSF applied remains fixed, it is more appropriate to phrase this in terms of the reduction of the loop's minor radius, $R$ (or the inferred value $R_L$), resulting in a lower number of points across the loop's transverse intensity profile. The normalised density contrast in the density profile model, $A$, cannot be compared to the actual density contrast so it's dependence on $R_L$ is not included. In Figure \ref{res_test} the inferred value of $\epsilon$ and its credible interval are plotted as a function of $R_L$, for seven different values of \lq downsampling\rq~of the data. The fitted value of $\epsilon$ from the actual density profile is overplotted (dashed line). There is a systematic negative offset from the actual value of $\epsilon$, which is expected due to the isothermal approximation made in the forward modelling method. In the context of real observations this offset is negligible, however. The bigger effect is the divergence which occurs for the two lowest values of $R_L$ tested. Unfortunately the uncertainties obtained are not large enough to account for the offset. The important factor in the accuracy of the method is the number of data points across the inhomogeneous layer, the small inhomogeneous layer used represents the lower limit of where this method is applicable.

In Figure \ref{res_test} the two vertical dotted lines correspond to the chosen resolutions R1 and R2, where $R_L$ = 29  and 7, or $R$ = 32 and 8. This offset between the inferred and actual minor radius of the loop comes from the isothermal approximation made in the forward modelling. R1 tests future high resolution data, or unrealistically wide loops observed with AIA, R2 represents AIA resolution when observing sufficiently wide coronal loops with a fitted radius (of the density profile) of $ R_{L} \approx$ 10 pixels (4.5 Mm) \citep[see][]{2017A&A...600L...7P}.

\subsection{Noise test}
The effect of varying $A_N$, which determines the amplitude of the noise $N$ in Equation~2 was tested. There is little effect on the inferred values of $\epsilon$,  $A$ and $R_{L}$, but a large increase in the uncertainty (95\% credible interval). This increase in the uncertainty was greater for $\epsilon$, showing that the details of the TDP may become masked to the method used at higher noise levels. Loops with larger inhomogeneous layers would be less affected by this.


\begin{figure*}
\centering
\includegraphics[width=8.9cm]{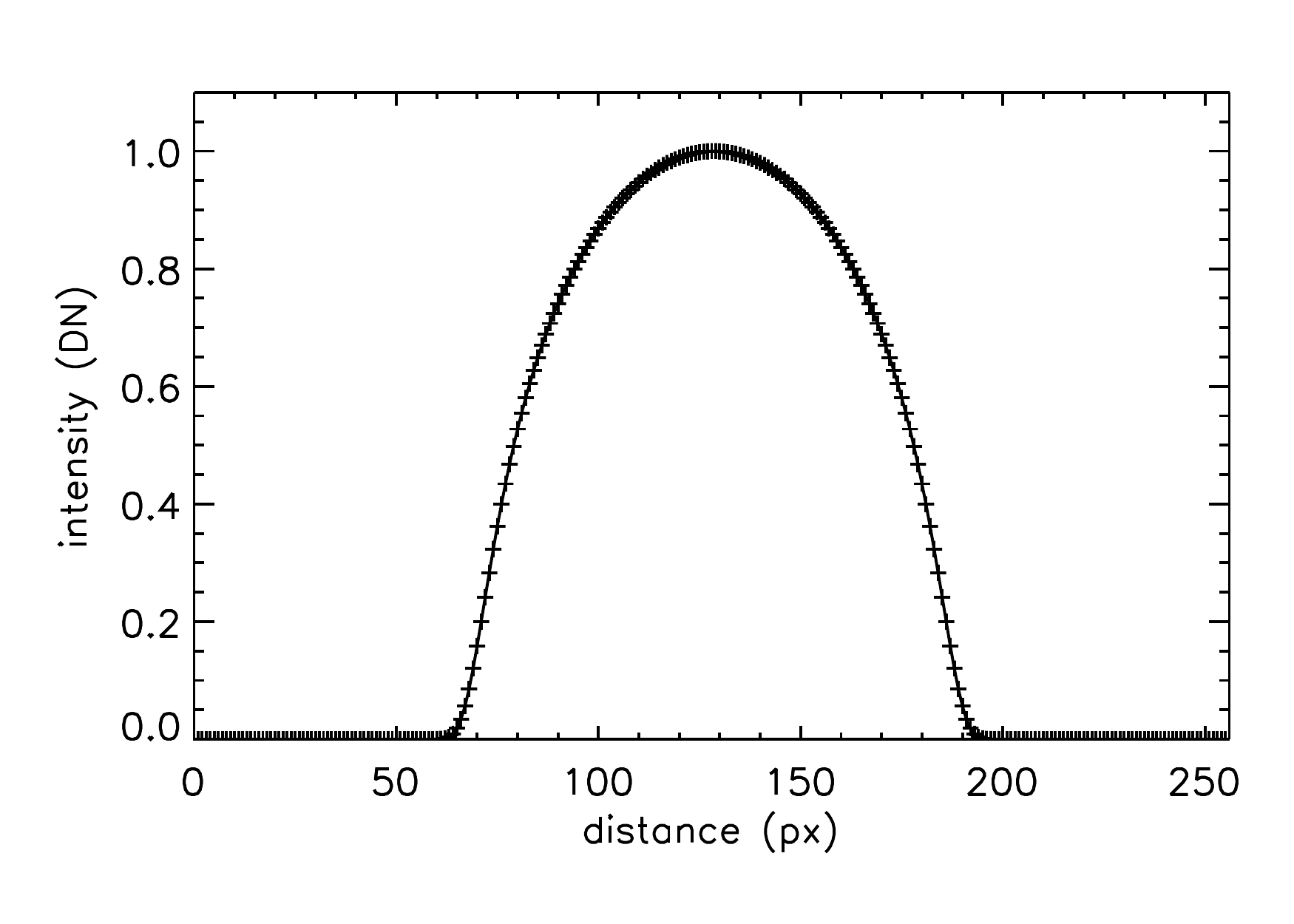}
\includegraphics[width=8.9cm]{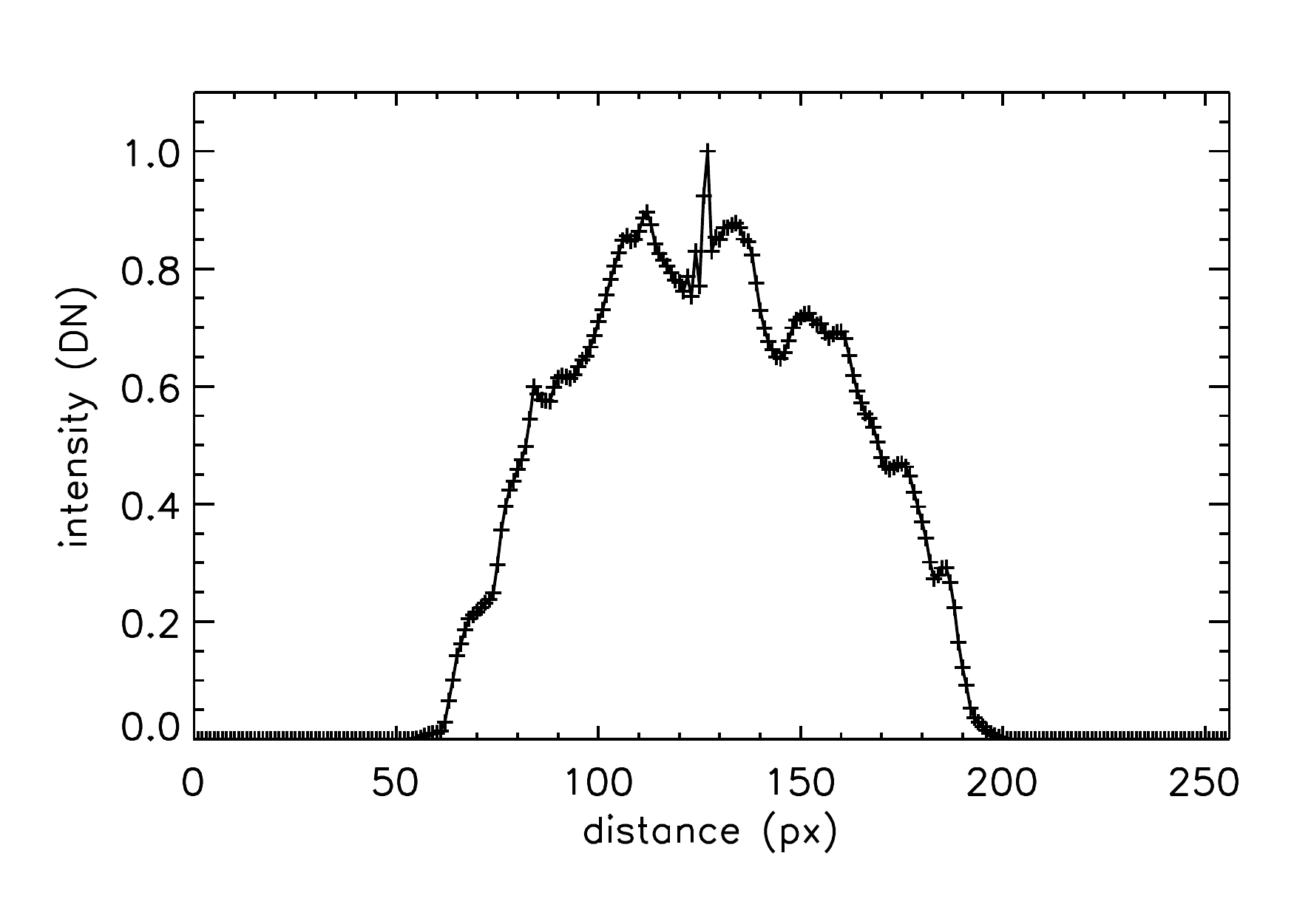}
\includegraphics[width=8.9cm]{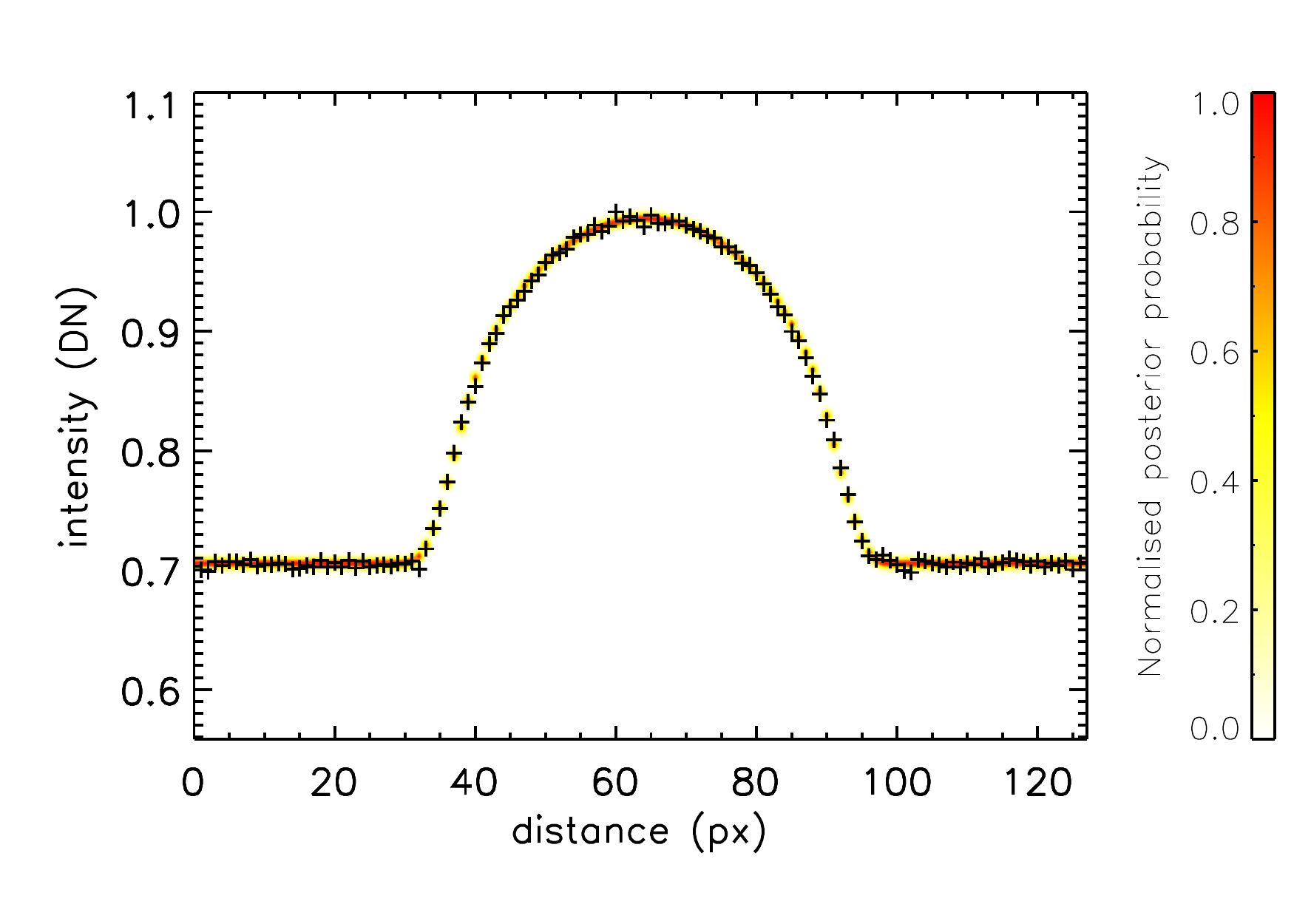}
\includegraphics[width=8.9cm]{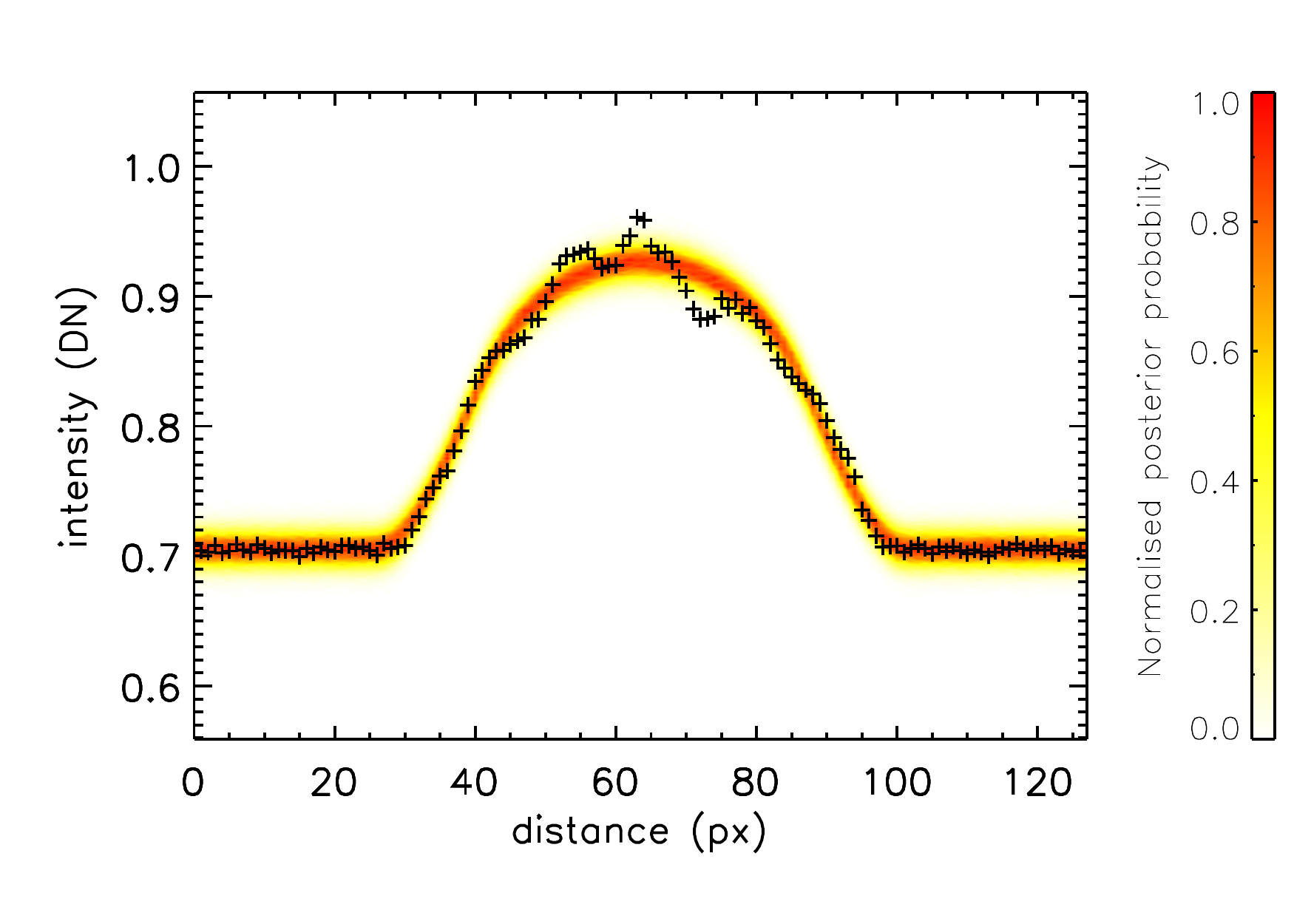}
\includegraphics[width=8.9cm]{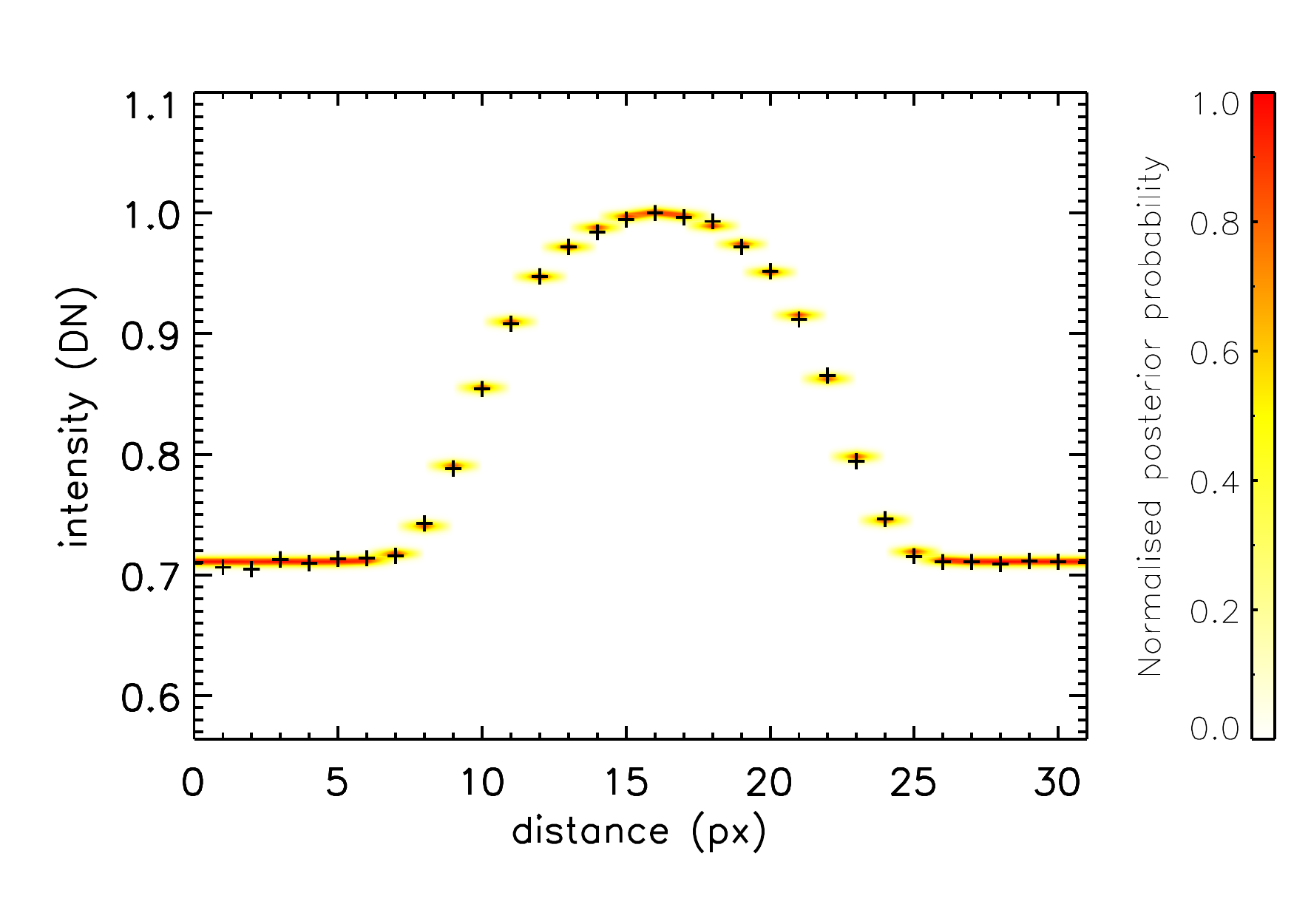}
\includegraphics[width=8.9cm]{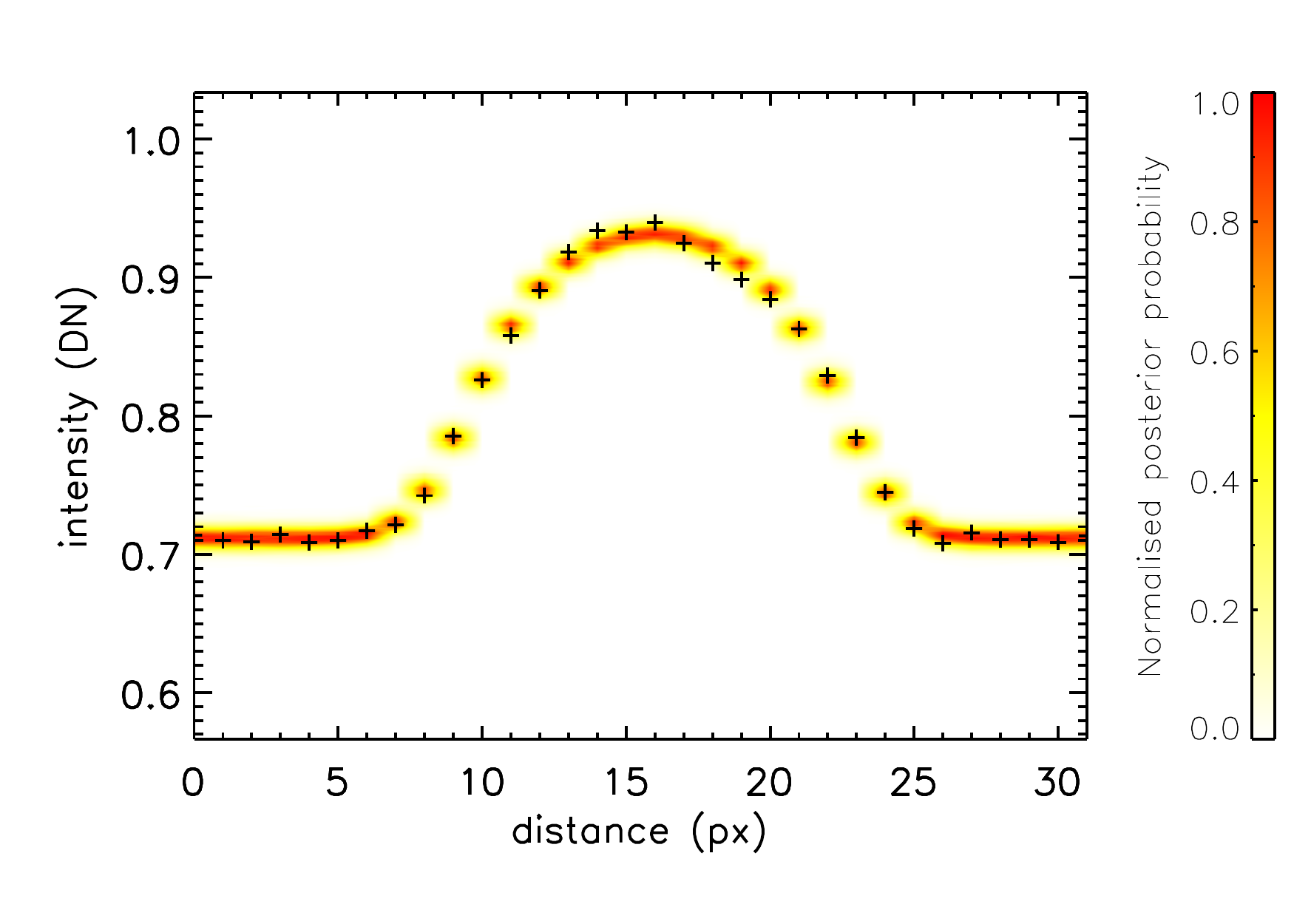}
\caption{Top row: Initial transverse loop intensity profile for M1 at the original resolution of the simulation, and the final intensity profile. Middle row: The initial and final intensity profiles at R1 with simulated noise and the PSF applied and the normalised posterior probability for each point plotted in the background. Bottom row: as above but for R2.}
\label{m1_profs}
\end{figure*}

\begin{figure*}
\centering
\includegraphics[width=8.9cm]{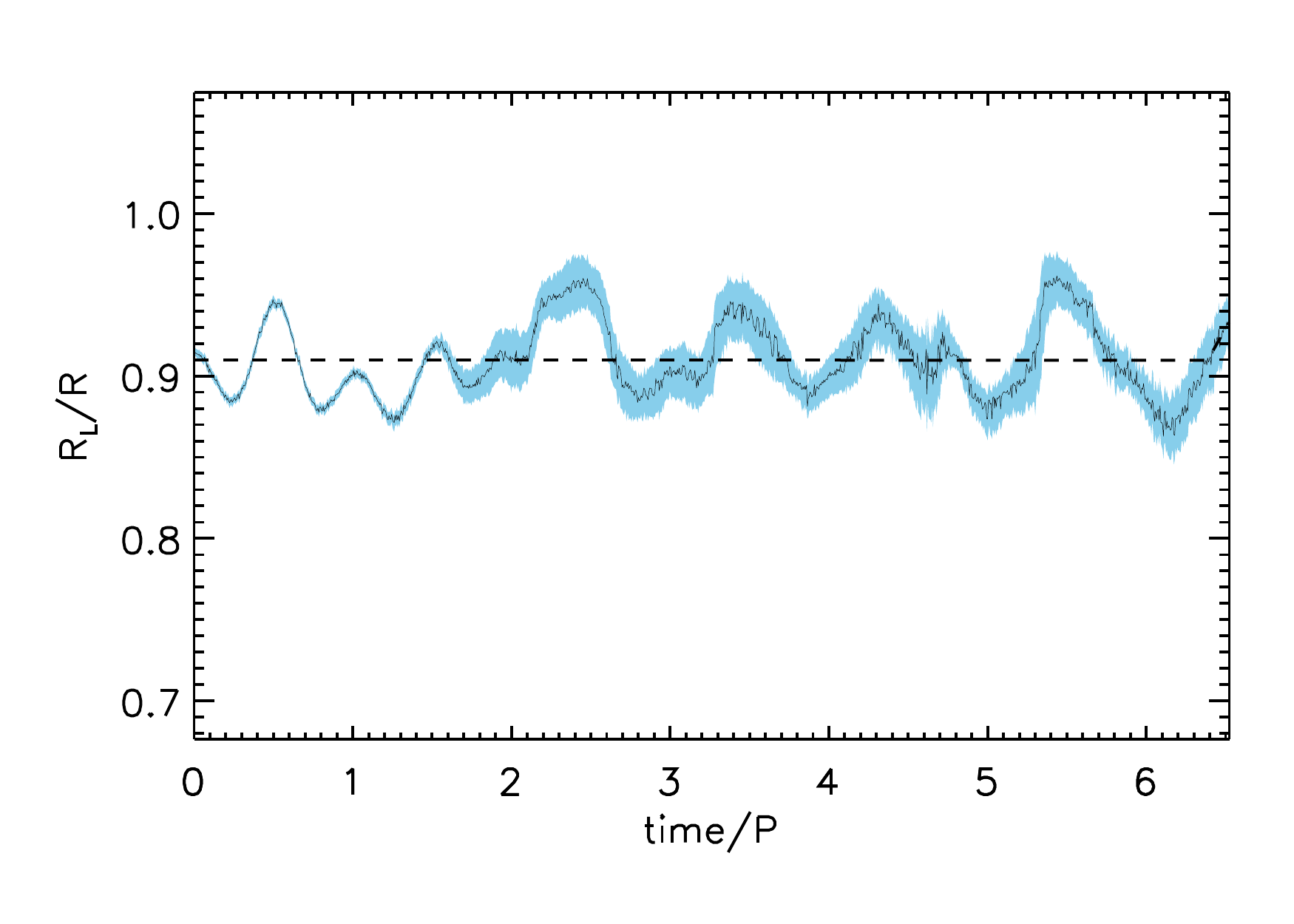} 
\includegraphics[width=8.9cm]{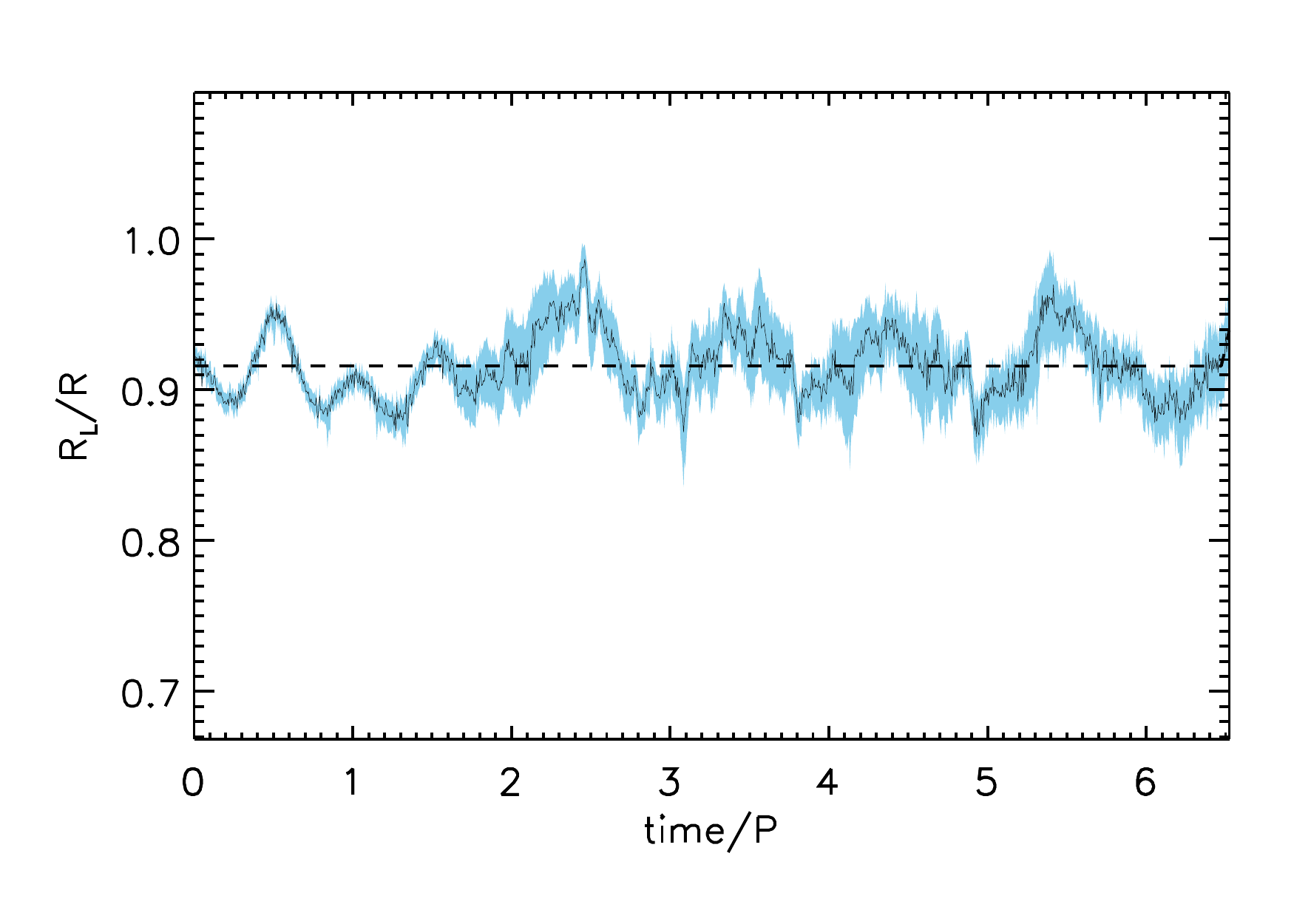} 
\includegraphics[width=8.9cm]{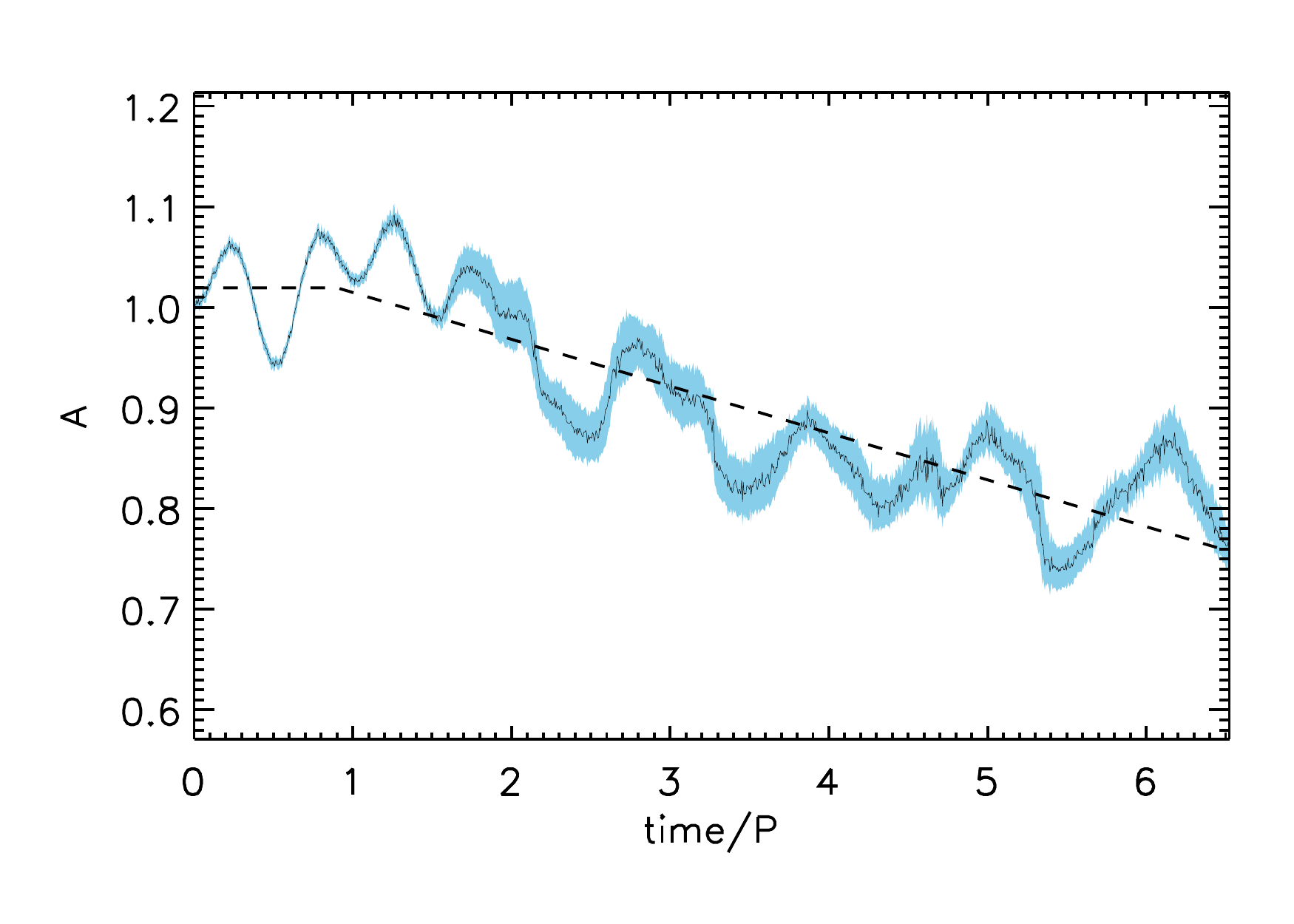}
\includegraphics[width=8.9cm]{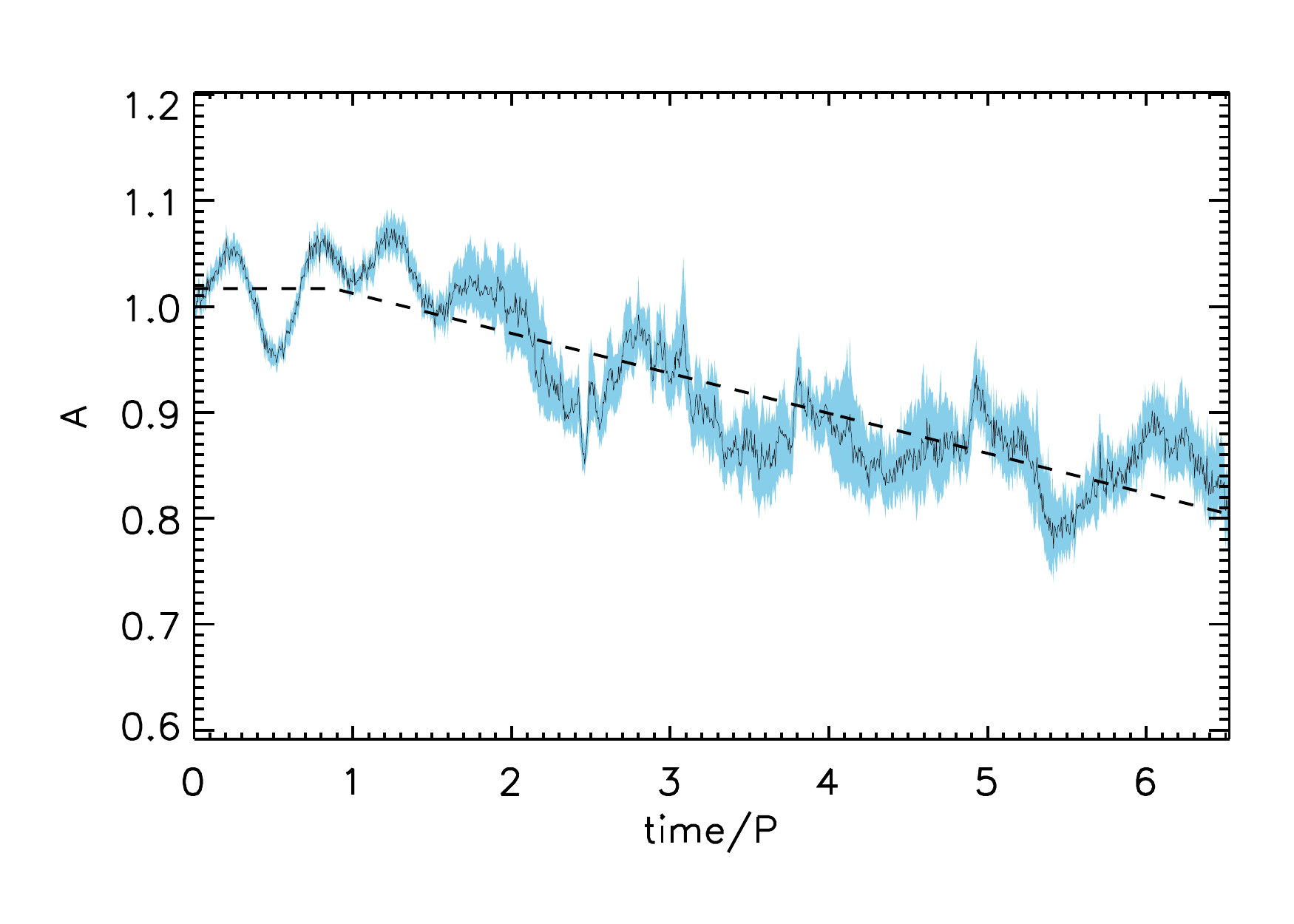}
\includegraphics[width=8.9cm]{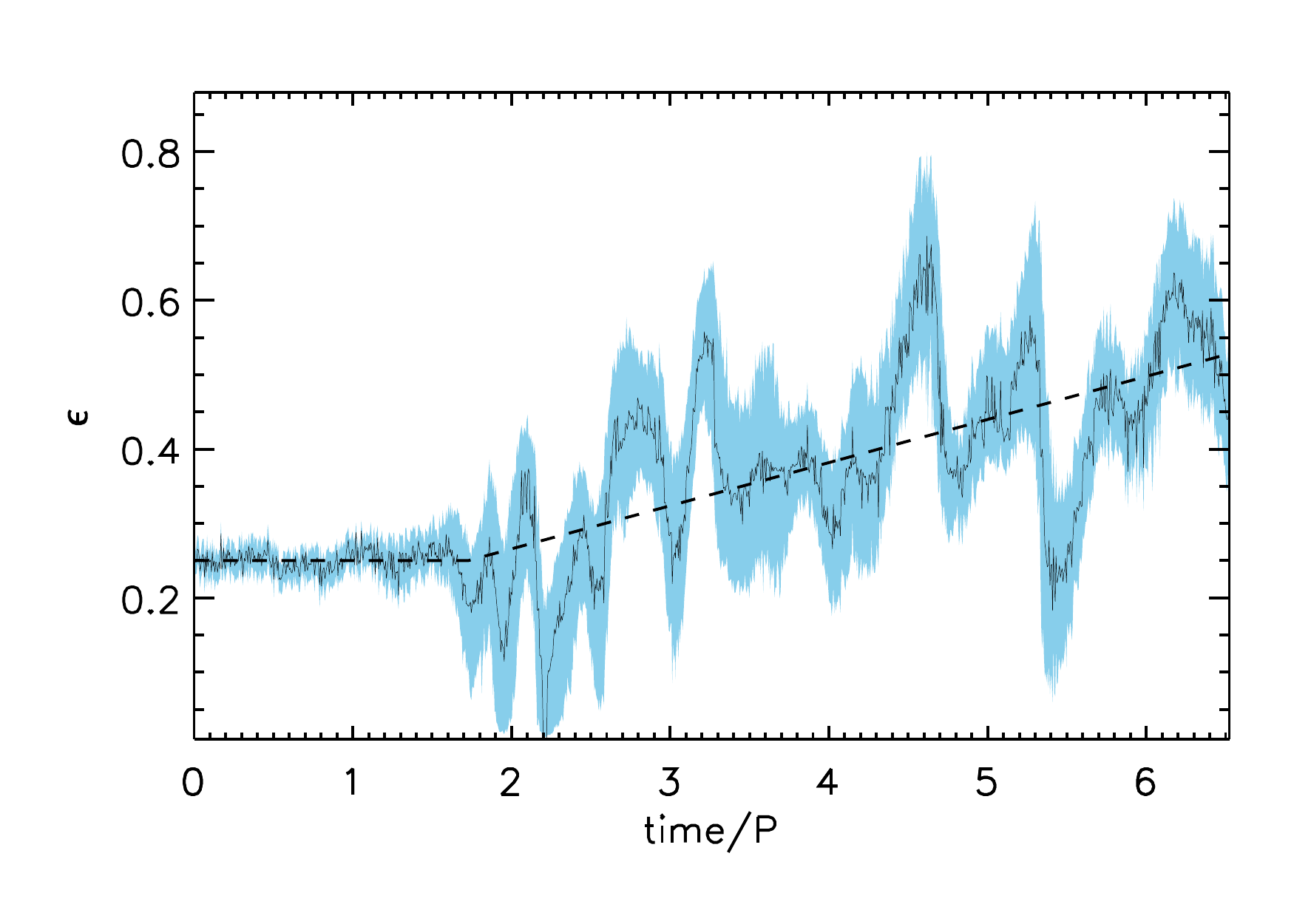}
\includegraphics[width=8.9cm]{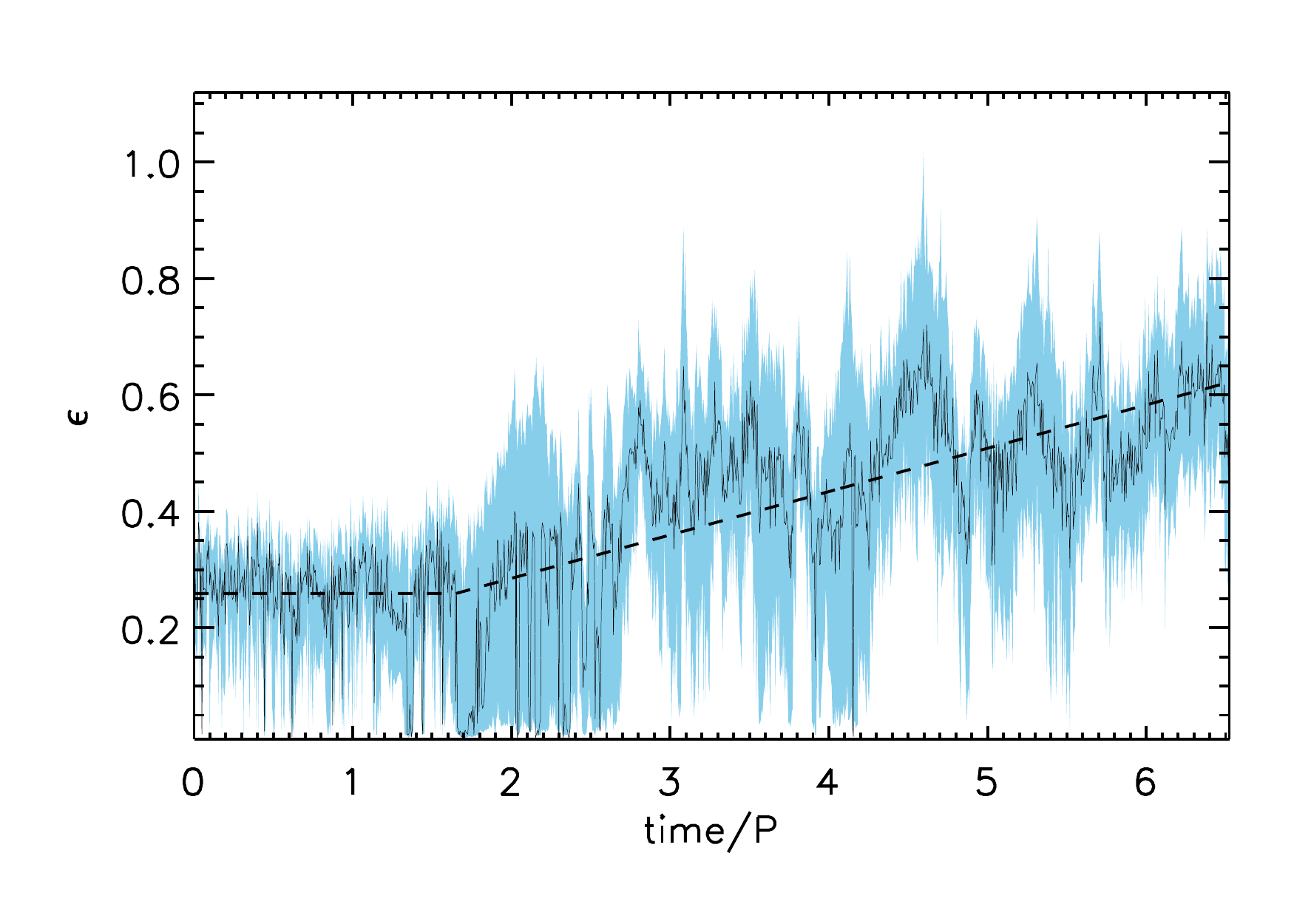}    
\caption{Time series of MAP values of $R_{L}/R$ (the inferred loop radius normalised by the actual radius), normalised $A$ (the normalised density enhancement) and $\epsilon$ (the inhomogenous layer width) and their uncertainty (shaded blue region) inferred from the M1 TD map with the simulated noise and PSF added. The uncertainty corresponds to the upper and lower values of the 95 $\%$ confidence interval. The left column corresponds to resolution R1 ($R$ = 32 pixels) and the right column to a four times lower resolution, R2 ($R$ = 8 pixels).}
\label{m1_ts}
\end{figure*}

\section{Time evolution - M1}\label{sec:m1}

The time distance maps for M1 at R1 and R2 are shown in Figure \ref{m1_tdmaps}. The added noise, reduced spatial resolution and Gaussian blur from the PSF mask the effect of the TWIKH rolls in the TD maps, particularly in the later case.  Hence it is important to determine what signatures of the TDP evolution of the loop are obtained when the TWIKH rolls cannot be resolved. Even the lower resolution R2 is actually a rare scenario in current EUV observations in terms of the number of points across the loop, as loops this wide are normally made up of a small number of visible strands/threads.

\subsection{Snapshot comparison}
The intensity profile for the initial and final frames of the M1 TD map are compared. In observations, the loop may not be in equilibrium, and as such any long term variation would have to be accounted for to obtain the evolution due to the oscillation only. The initial and final profiles for Model 1 at the full resolution of the simulation are plotted in the top two panels of Figure \ref{m1_profs}. The middle two panels are the same but with the added noise and resolution R1, and the bottom two panels are at R2. The first column highlights the effect the lower resolution and noise have on the initial intensity profiles of the loop. The second column shows how the intensity profile has been disrupted by the KHI vortices, and also how the noise and reduced resolution can mask the peaks in the intensity profile that appear. At the lower resolution the individual peaks are not resolved and the intensity profile merely appears slightly asymmetric. 

\subsection{Time series}

The above comparison is now extended to produce time series for the TDP parameters of interest. This is done by applying the forward modelling approach to each intensity profile from the TD maps. 

In the left panels of Figure \ref{m1_ts} the time series for $A$, $R_{L}/R$ and $\epsilon$ are plotted for M1  at R1. The normalisation factor $R$ is the actual minor radius at the beginning of the simulation, and does not change over time. The overplotted dashed lines correspond to fits of the trends, which are linear fits with two sections, one with a gradient of zero and one with the gradient as a free parameter. This allows the general long timescale behaviour of a given parameter during the oscillation to be approximated. Both $R_{L}$ and $A$ time series exhibit oscillations due to the disruption of the loop cross-section by the transverse waves and the induced TWIKH rolls. In addition to this, background trends are detected. In agreement with the results in \cite{2017ApJ...836..219A}, $A$ is found to decrease once the TWIKH rolls and turbulence begin at $t$  $\approx$ 2 $P$. However the increase of the loop width previously detected is not present here. This is due to our use of Model L for the density profile, which allows the density enhancement ($A$) and the width of the inhomogeneous layer ($\epsilon$) to be decoupled from the radius ($R_L$). Hence a large increase in the value of $\epsilon$ with time is detected, varying from 0.28 to 0.6 at the end of the time series.  

In the right panels of Figure \ref{m1_ts} the same data is analysed at R2. The inferred values are as above but with larger uncertainties, and more noise in the time series itself. This shows that the reduced spatial resolution does not change the inferred values of the TDP parameters significantly, as expected from the tests in Section 3.2. Some of the fine structure in the time series would have been reduced by integrating in time to a more realistic cadence, similar to that of AIA, meaning real observations may appear less noisy than the time series presented here. 

\begin{figure}
\centering
\includegraphics[width=8.9cm]{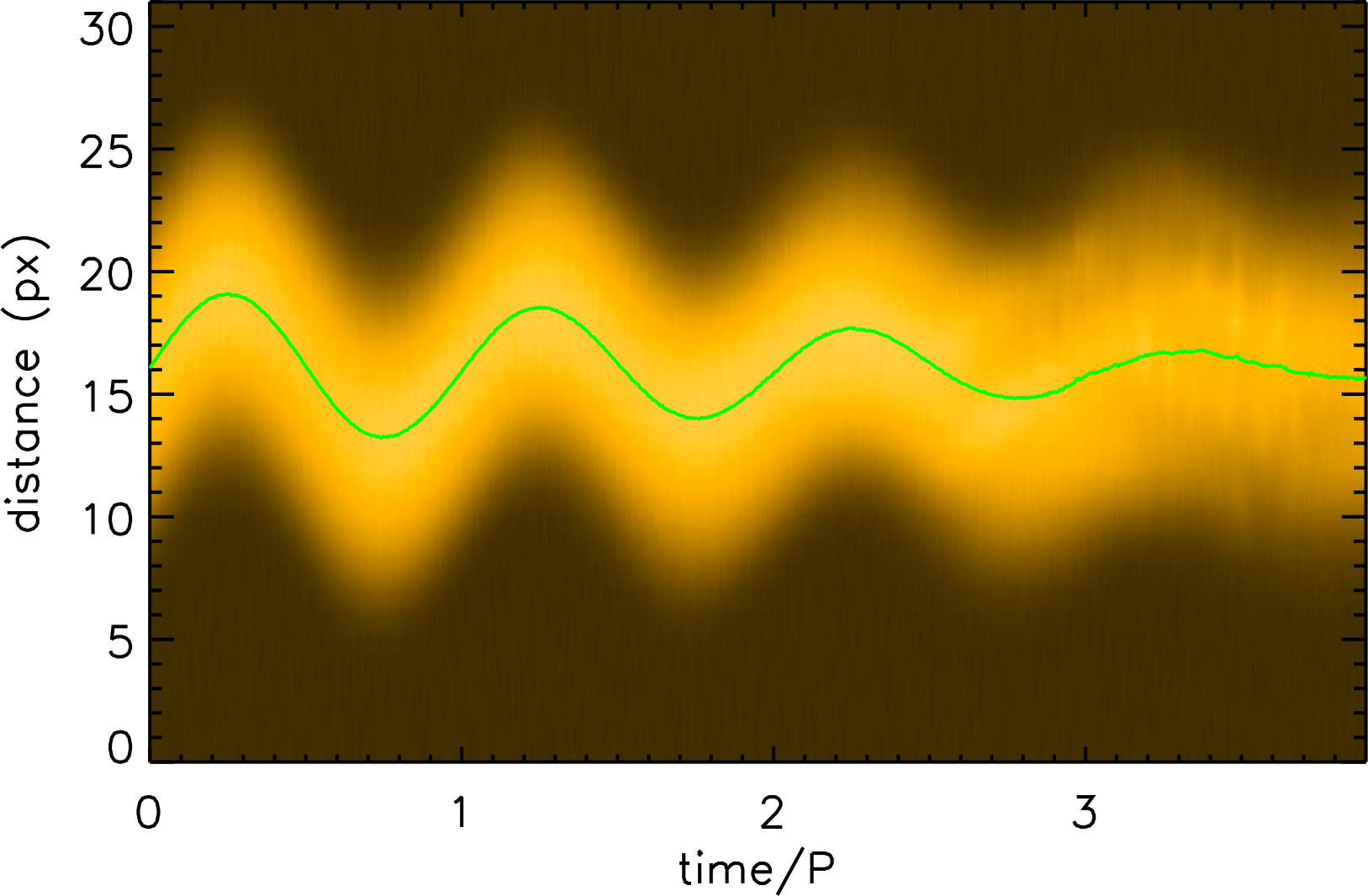}
\includegraphics[width=8.9cm]{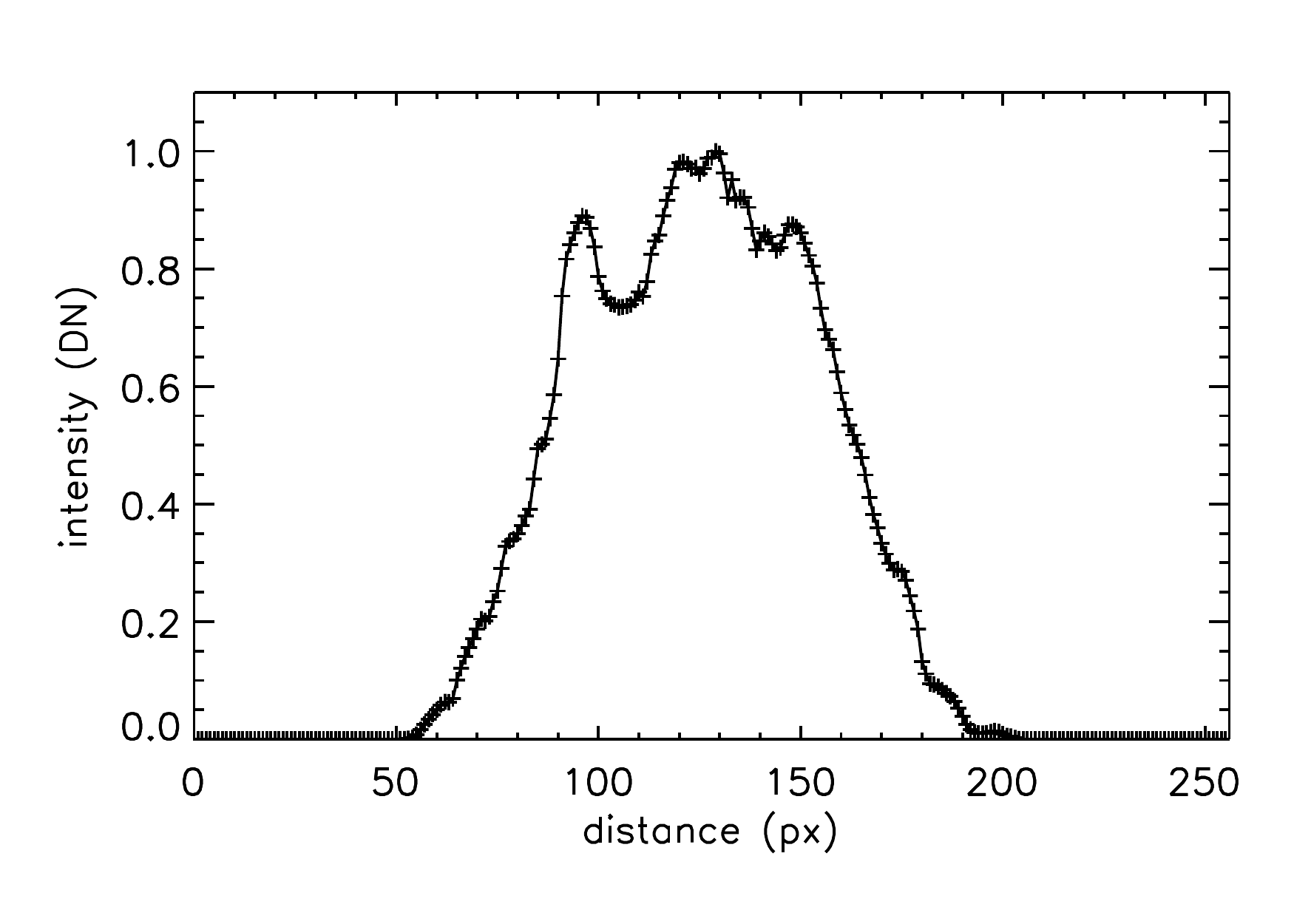}
\includegraphics[width=8.9cm]{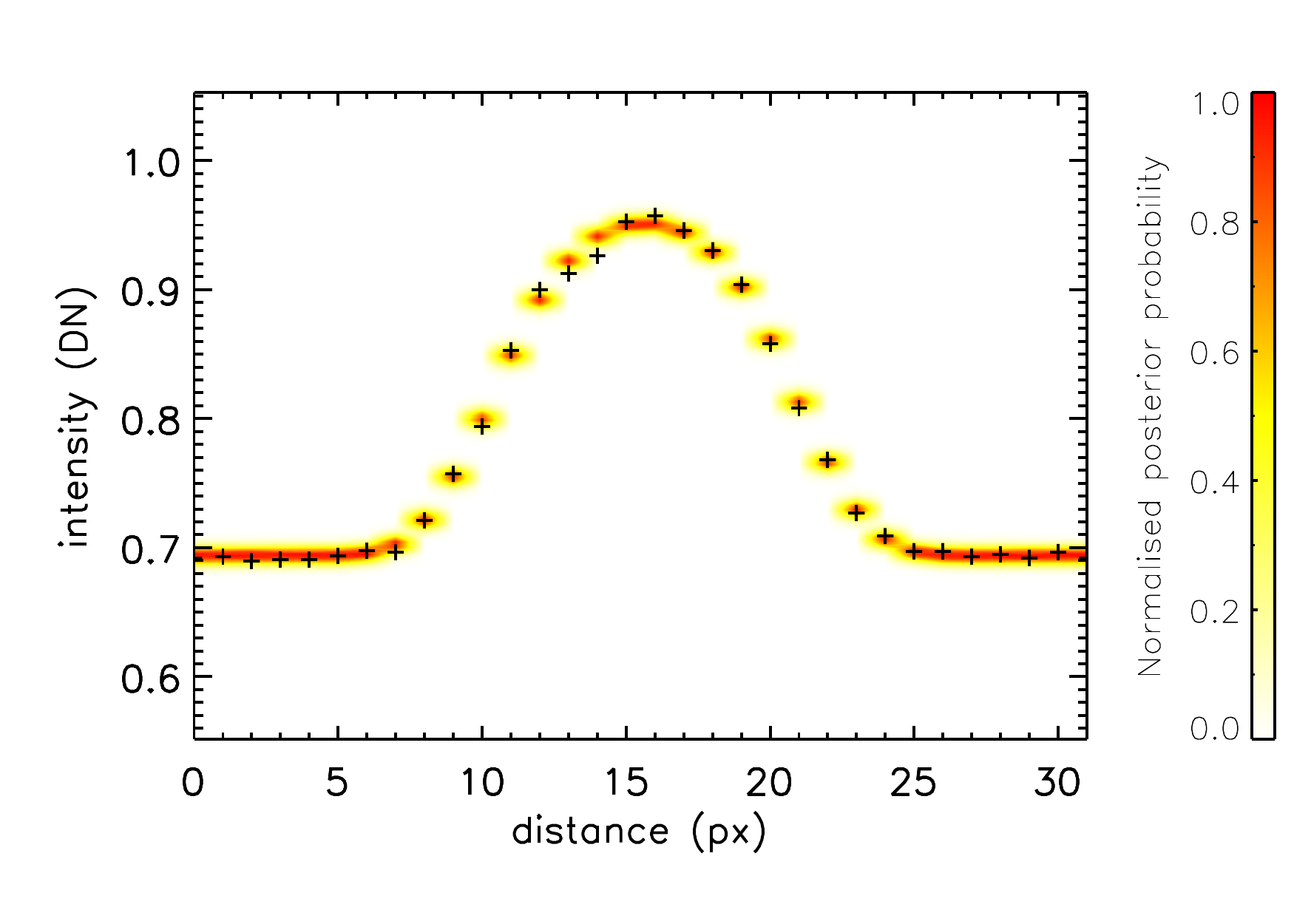}
\caption{Top: TD map for M2 at resolution R2. The overplotted green line corresponds to the MAP value of the loop centre position at each time. Middle and Bottom: final intensity profile for M2 at the original resolution and the same at R2 with simulated noise and the PSF applied and the normalised posterior probability for each point plotted in the background.}
\label{m2_profs}
\end{figure}

\begin{figure}
\centering
\includegraphics[width=8.9cm]{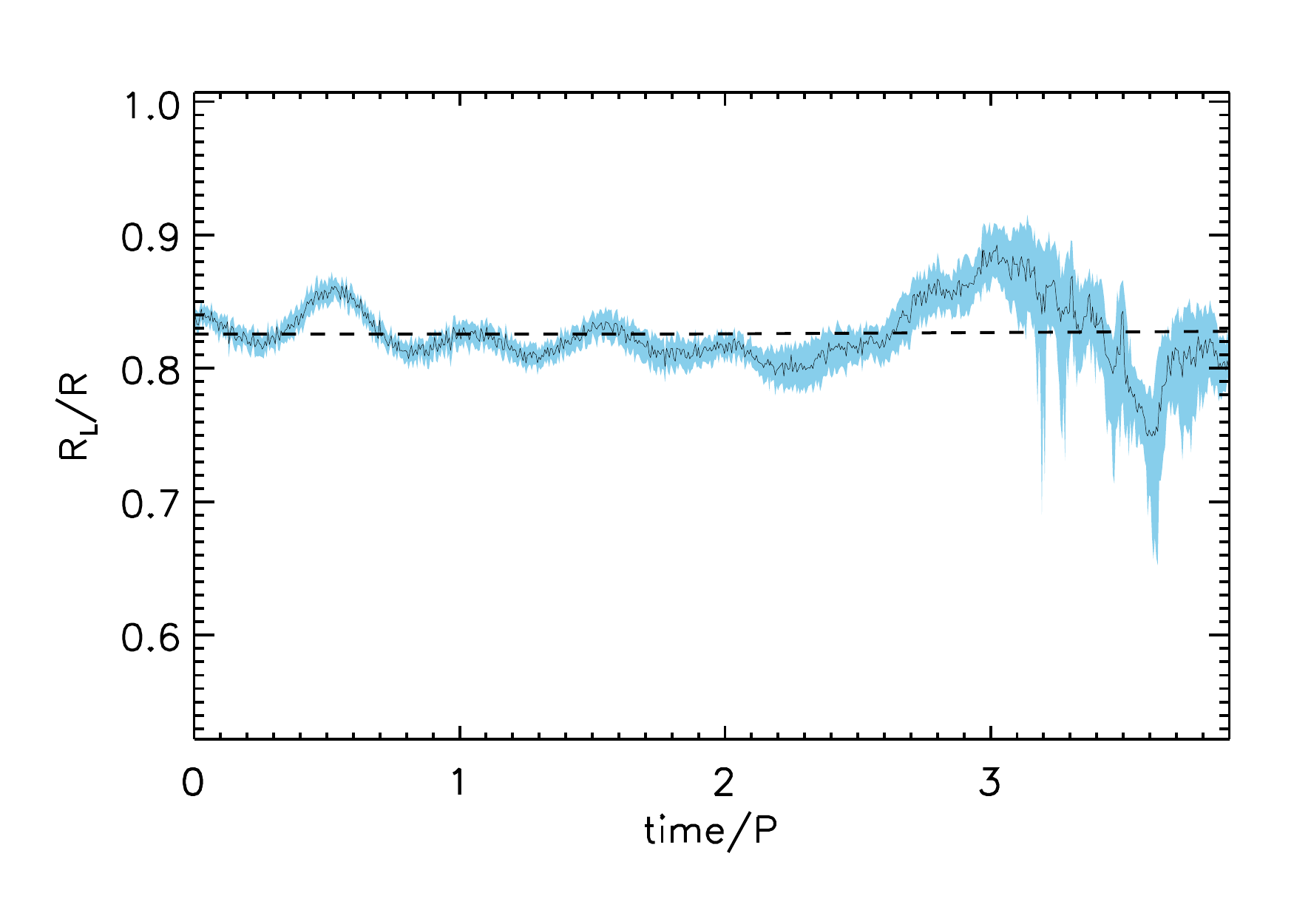} 
\includegraphics[width=8.9cm]{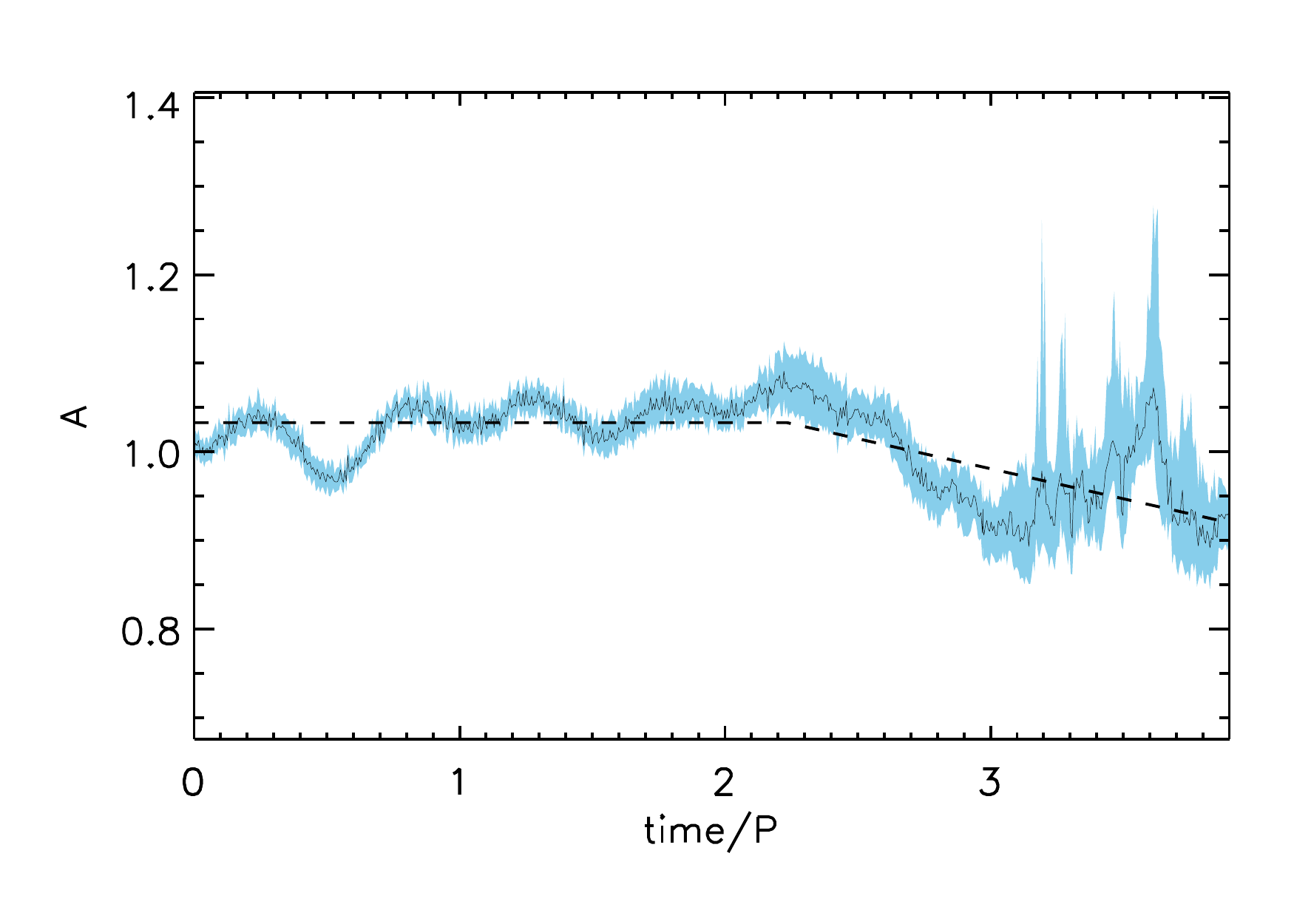}
\includegraphics[width=8.9cm]{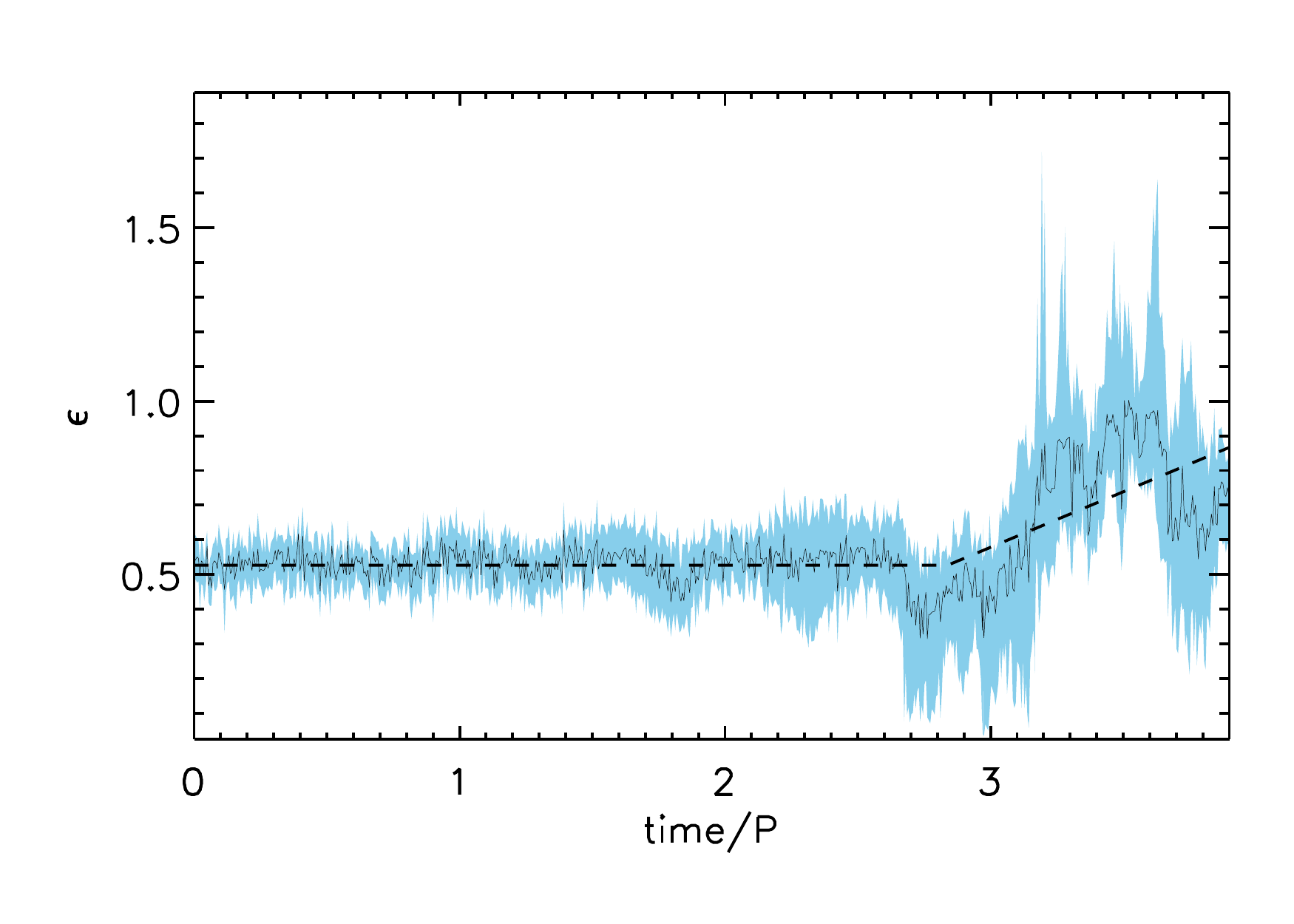}     
\caption{Time series of MAP values of $R_{L}/R$, normalised $A$ and $\epsilon$ (black points) and their uncertainty (shaded blue region) inferred from the M2 TD map at R2. }
\label{m2_ts}
\end{figure}

\section{Time evolution - M2 \& M3}\label{sec:m2m3}
\subsection{M2} 
Analysing M2 allows the effect of a larger inhomogeneous layer to be compared to the results obtained for M1. The TD map for M2 at R2 is shown in Figure \ref{m2_profs}. It can be seen that the effects of the TWIKH rolls are delayed until $t$ $\approx$ 3 $P$. This was discussed in Section \ref{init_td}. An exploration of the effect of the width of the inhomogeneous layer on the development of the KHI instability was made in \cite{2016A&A...595A..81M}. Once the KHI is generated visual inspection of the TD map shows a drop in intensity as well as an apparent broadening of the loop. Some high frequency time variation is apparent, as in M1 at R2, due to out of phase TWIKH rolls being integrated together spatially. In observations the time span would be extended beyond the TD map shown here, and stronger disruption of the loop should be evident.

The final intensity profile for Model 2 at the full resolution is plotted in the second panel of Figure \ref{m2_profs}. Sharper intensity (and therefore temperature and density) gradients are formed compared to M1, despite the onset of the KHI being delayed. In the bottom panel it is shown at R2. At the lower resolution the individual peaks are not resolved and the intensity profile merely appears slightly asymmetric, as seen for M1.
 
The time series of the density profile parameters for model M2 (Figure \ref{m2_ts}) exhibit similar oscillations to M1. This shows that this effect is largely due to the transverse oscillation, as the TWIKH rolls develop later in the case of M2. After KHI onset the parameters increase and decrease in similar ways, however this effect is delayed until $t$ $\approx$ 3 $P$, shortly before the end of the time series. The increase in the inferred value of $\epsilon$ is large, increasing from 0.5 to 1.0 at the end of the time series. $A$, the normalised density enhancement, decreases by $\approx$ 15 \%. The radius $R_L$ is caused to vary, but not in a systematic manner. In observations, which sometimes display more cycles of oscillation than are analyse here \citep[see][]{2016A&A...585A.137G}, and the dominance of large inhomogeneous layers found in \cite{2017A&A...605A..65G}, it can be postulated that the observational signature shown here should often be stronger in observations. 

\subsection{M3}

\begin{figure}
\centering
\includegraphics[width=8.9cm]{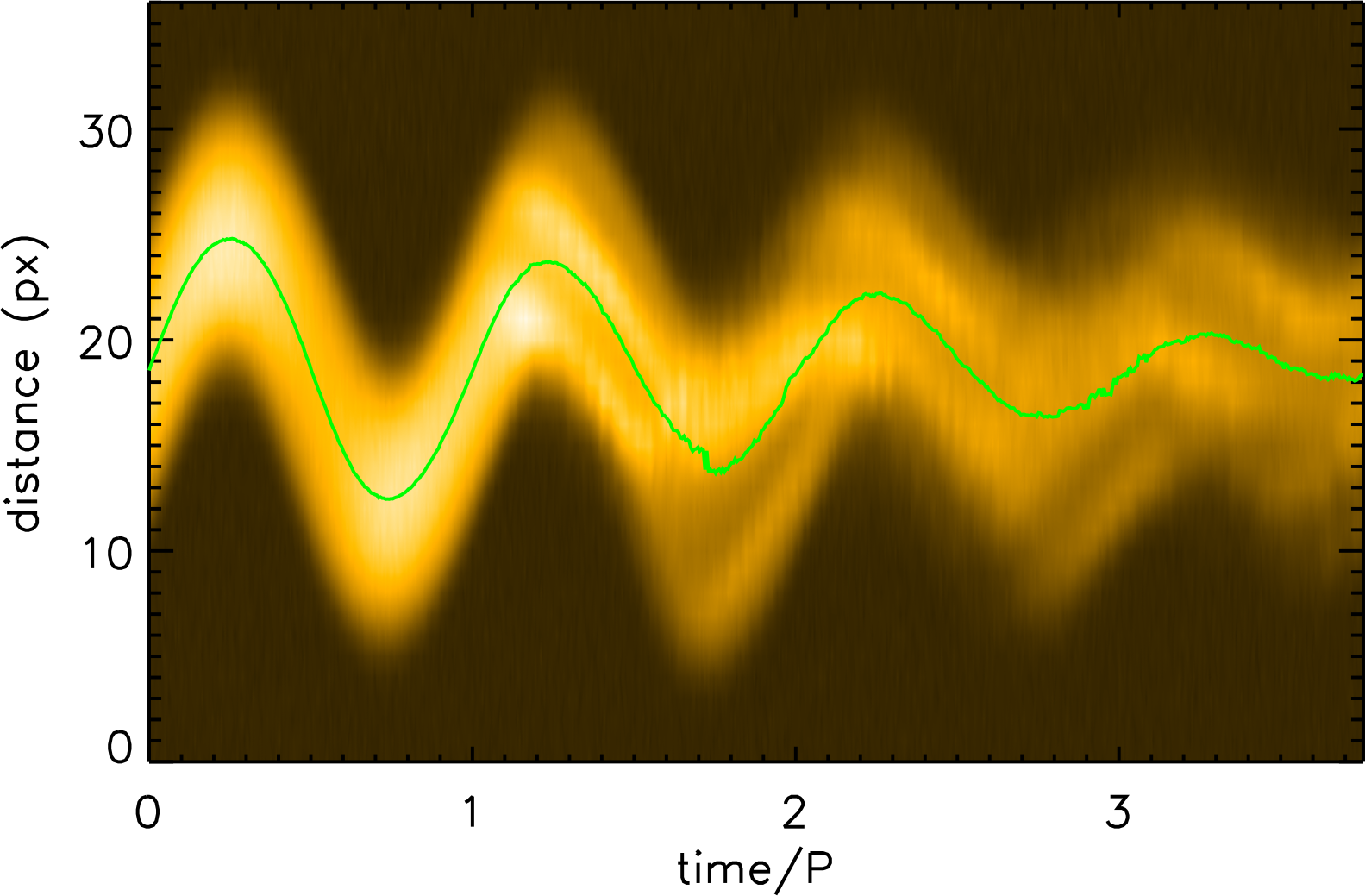}
\includegraphics[width=8.9cm]{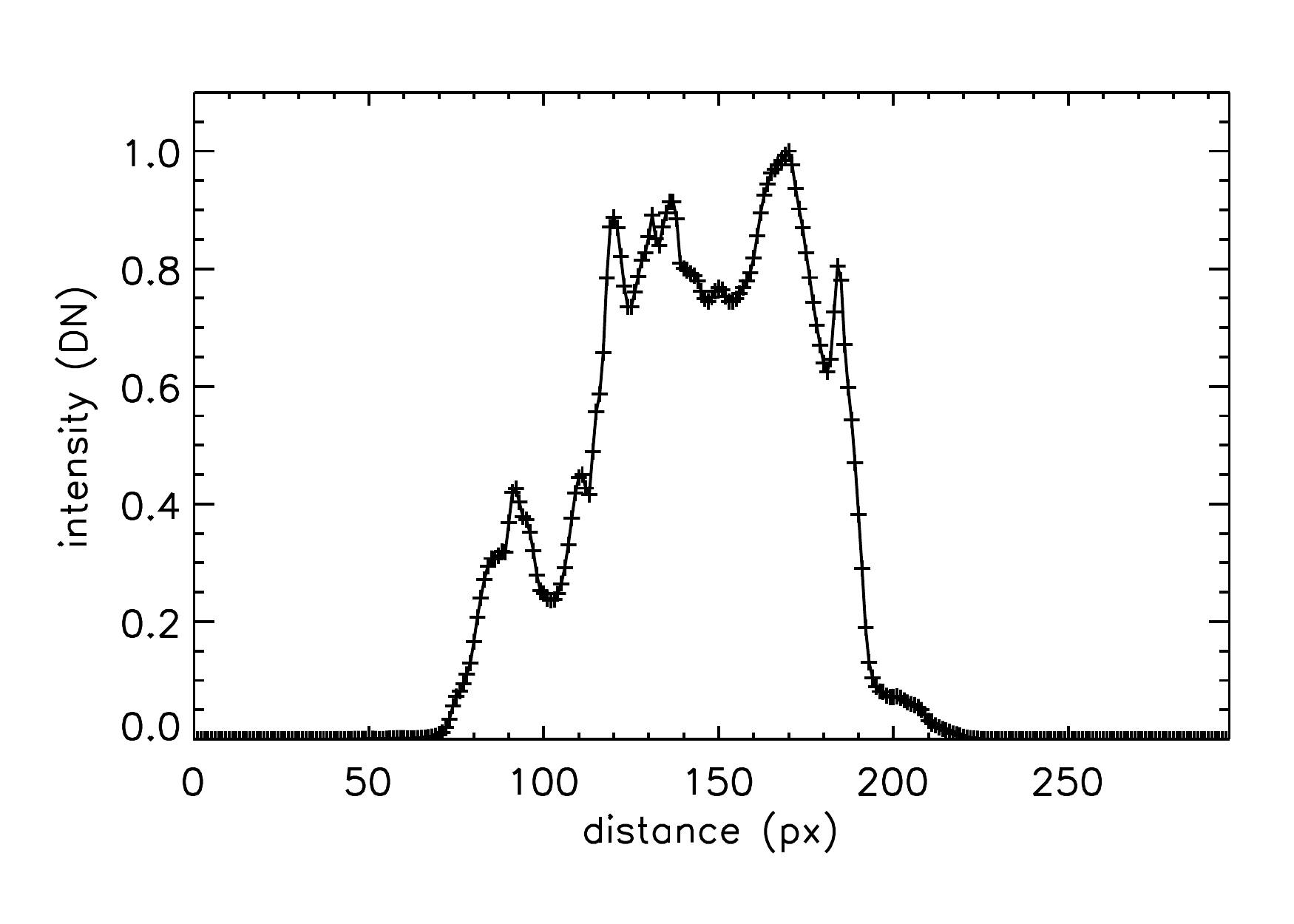}
\includegraphics[width=8.9cm]{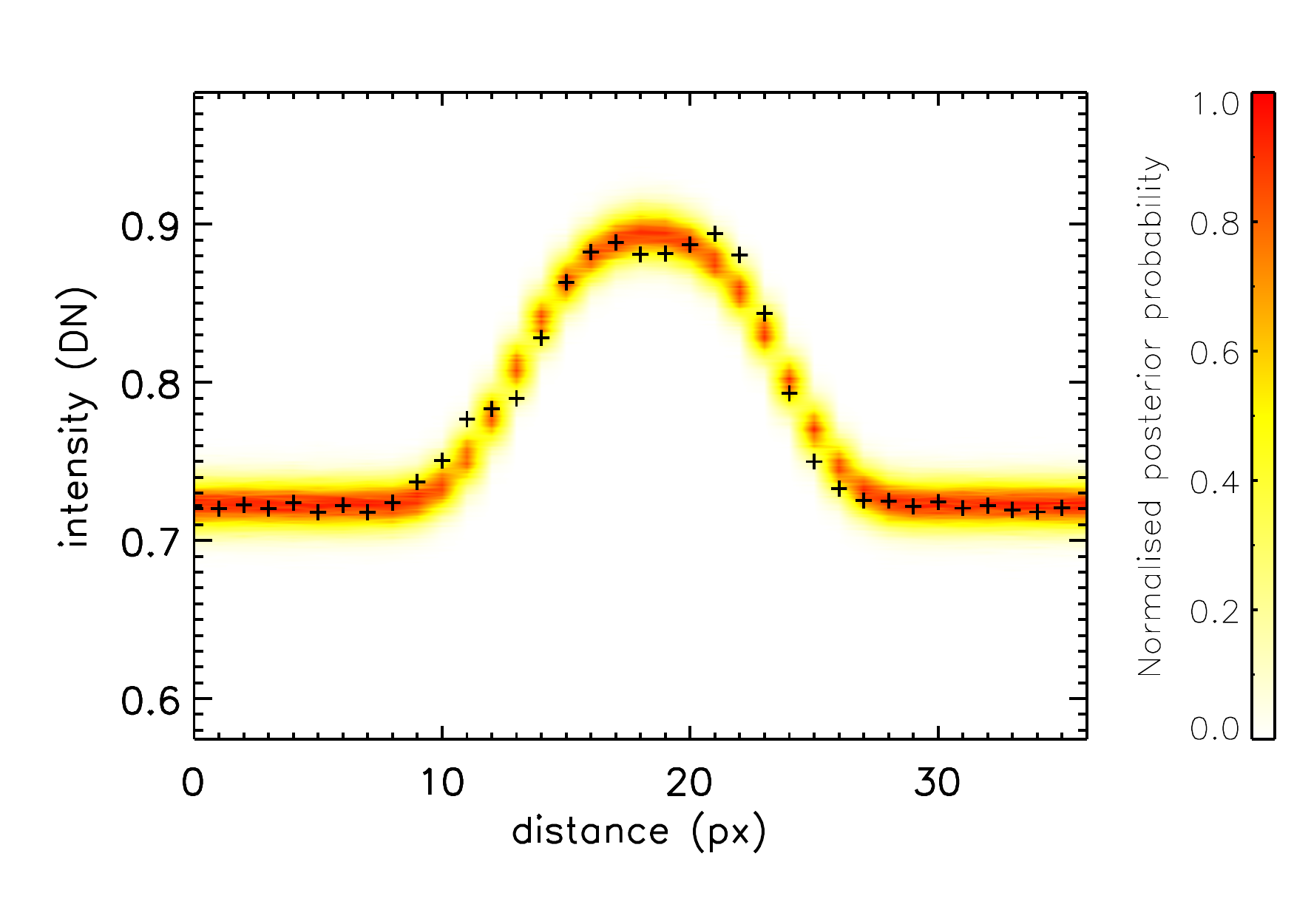} 
\caption{Top: TD map for M3 at resolution R2. The overplotted green line corresponds to the MAP value of the loop centre position at each time. Middle and bottom: final intensity profile for M3 at the original resolution and the same at R2 with simulated noise and the PSF applied and the normalised posterior probability for each point plotted in the background.}
\label{m3_profs}
\end{figure}

\begin{figure}
\centering
\includegraphics[width=8.9cm]{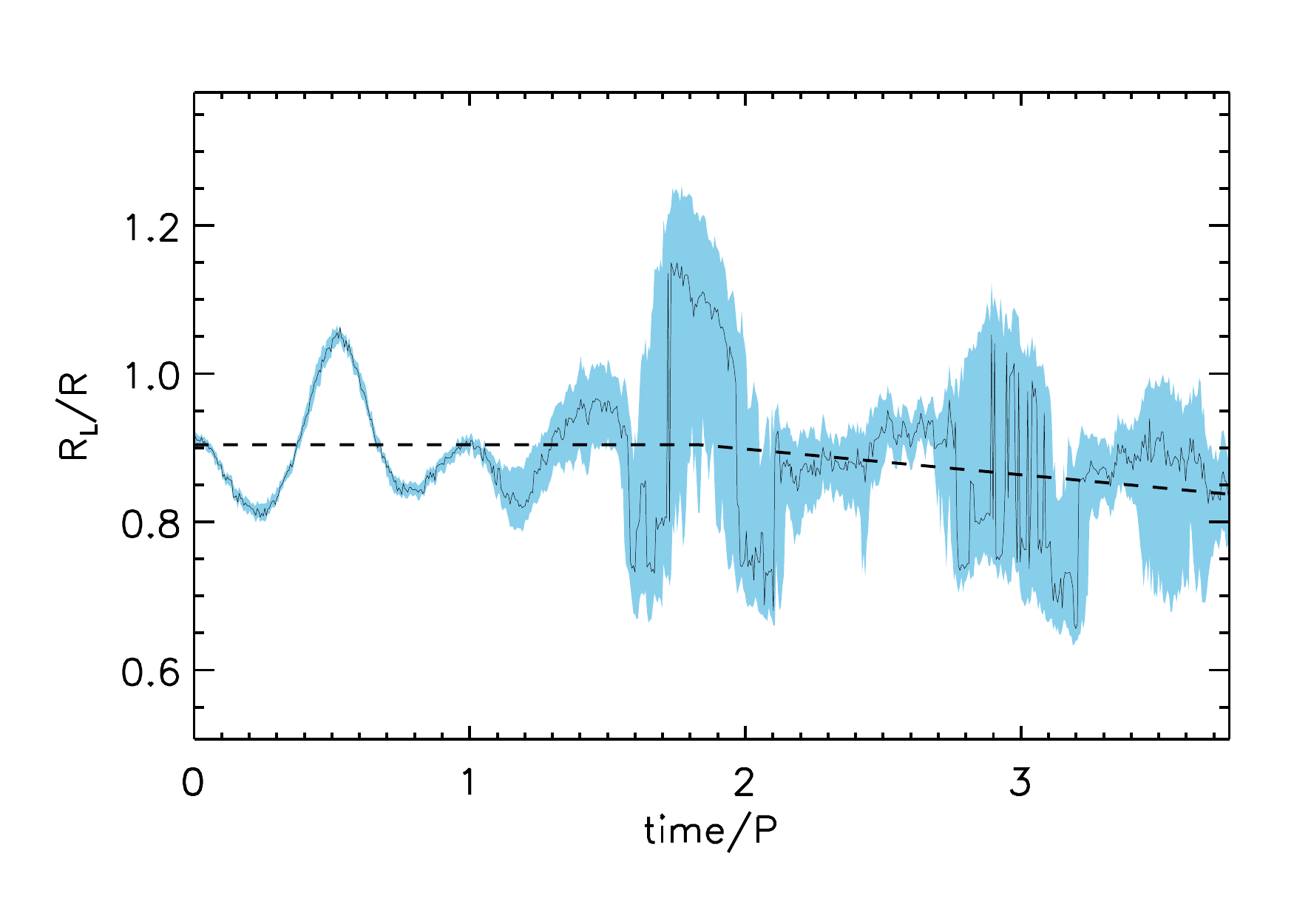} 
\includegraphics[width=8.9cm]{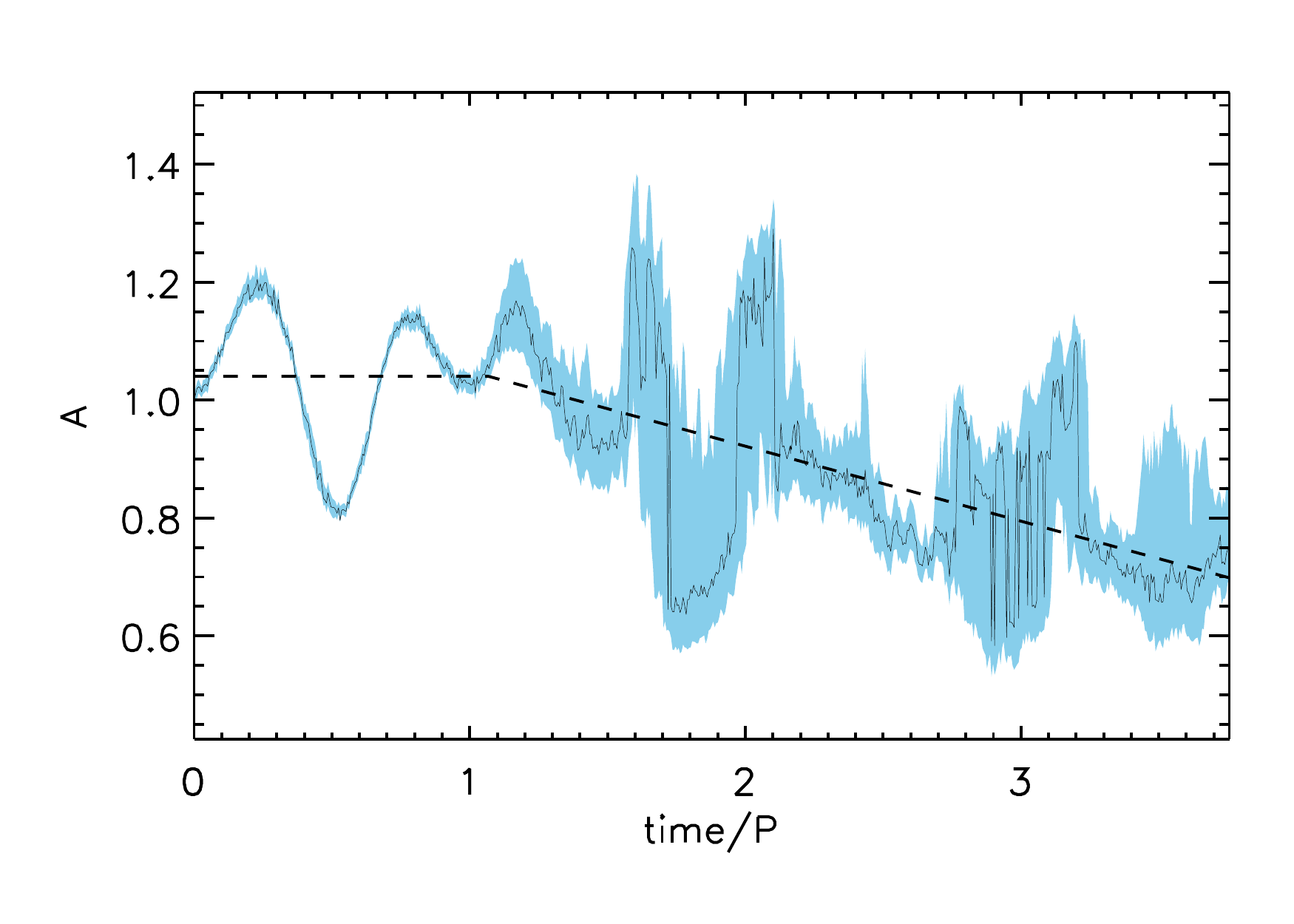}
\includegraphics[width=8.9cm]{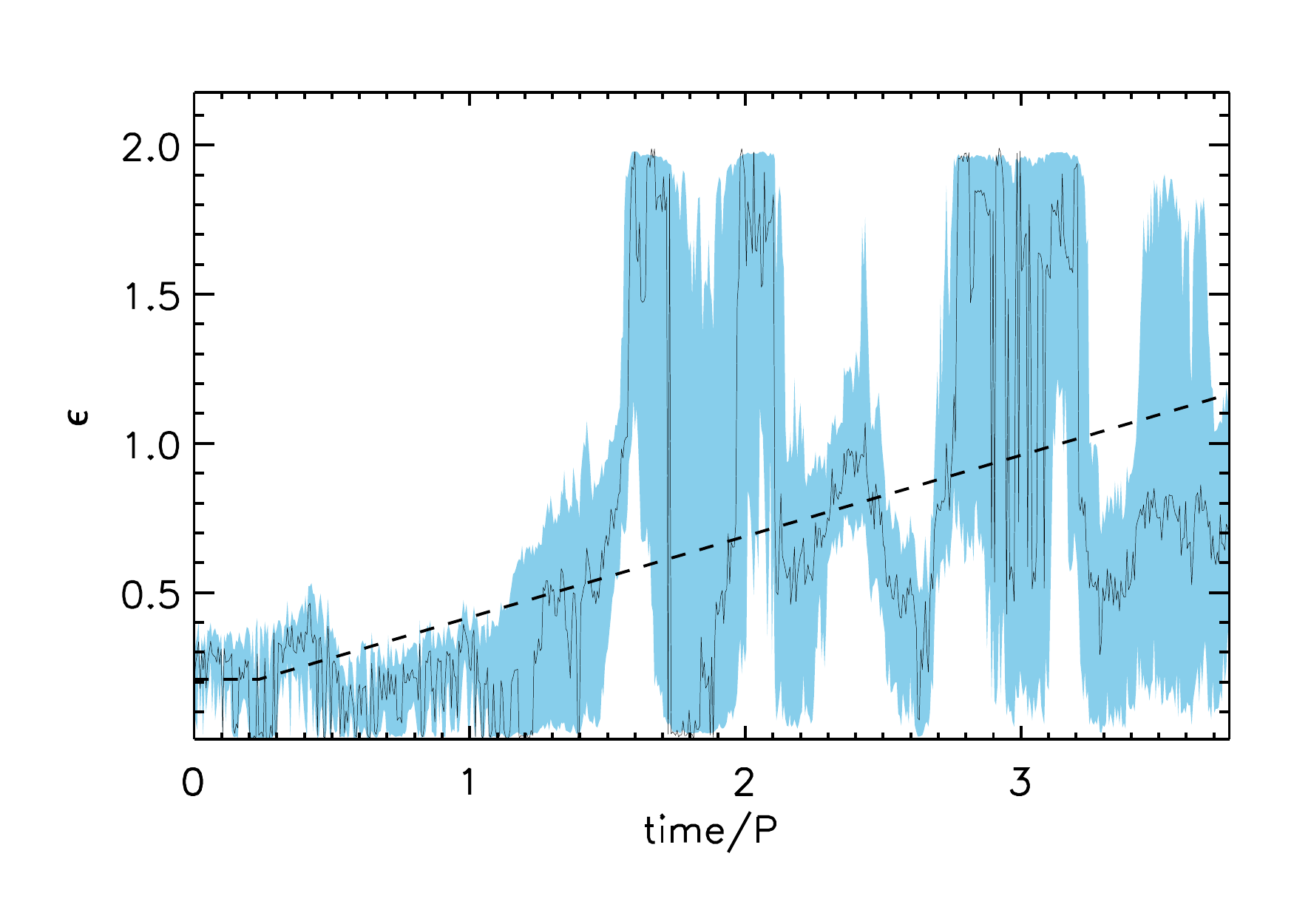} 
\caption{Time series of MAP values of $R_{L}/R$, normalised $A$ and $\epsilon$ (black points) and their uncertainty (shaded blue region) inferred from the M3 TD map at R2.}
\label{m3_ts}
\end{figure} 

Analysing M3 allows the effect of a larger oscillation amplitude to be compared to the results obtained for M1 and M2. The TD map for M3 at R2 is shown in Figure \ref{m3_profs}. The effects of the TWIKH rolls appear at $t$ $\approx$ 1 $P$. Some high frequency time variation is apparent, as in M1 and M2, due to out of phase TWIKH rolls being integrated together spatially. The disruption of the loop by the KHI is clear, even at the resolution R2. From visual inspection of the TD map there is a strong reduction of the intensity, and disruption of the transverse structure. The many TWIKH rolls visible in Fig. \ref{m1_tdmaps} appear as two or three slightly out of phase larger strands within the loop. 

The initial and final profile for M3 at the full resolution are plotted in the top panels of Figure \ref{m3_profs}. From the bottom left panel it is clear that the higher oscillation amplitude has created the strongest transverse intensity variation across the loop, and the sharpest gradients. In the bottom right the individual peaks are just resolved at R2 and the intensity profile is starting to become badly approximated by the TDP model, evident from the much more spread posterior probability for each data point. 
 
The time series of the density profile parameters for model M3 include higher amplitude oscillations than M1 and M2, due to the higher oscillation amplitude of the simulated loop. This again shows that this effect is due to the transverse oscillation. In addition the parameters vary in similar ways, however here this begins at after $t$ $\approx$ 1 $P$. A small decrease in the radius is detected, due to the intensity profile becoming badly modelled by the intensity forward modelled from the density profile. The increase in the inferred value of $\epsilon$ is large, increasing from 0.5 to 1.0 at the end of the time series. The decrease in the normalised density enhancement $A$ is also large, from 0.08 to 0.05, or 40 \%.

The credible intervals of the TDP parameters are much larger for $M3$, particularly for $\epsilon$, which becomes ill-constrained by the Bayesian inference during the development of the KHI. This is due to the strong intensity peaks from the TWIKH rolls not being fit by the forward modelled intensity profile. This does not pose an issue for the detection of the evolution of coronal loop TDPs, as the effects are clearly observed in the TD map itself in this case.

As mentioned above, large inhomogeneous layers are expected in a significant fraction of coronal loops, and oscillation amplitudes are often comparable to the amplitude in M3. It can be expected that if the KHI is generated as efficiently as in these simulations, strong evolution of the TDP of some oscillating coronal loops should be detected, even with current imaging instruments.
\section{Discussion}\label{sec:disc}

The main purpose of this study is to describe the inferred evolution of the loop's Transverse Density Profile (TDP) caused by non-linear effects which occur during kink oscillations (largely due to the KHI instability). This is motivated by recent advances in kink oscillation observations \citep[e.g][]{2016A&A...585A.137G}, modelling \citep[e.g][]{2017A&A...601A.107P, 2017A&A...602A..74H} and seismology \citep[e.g][]{2017A&A...603A.101L, 2017A&A...600A..78P}. For observational analysis and theoretical works which assume the transverse structure is stationary it is important to determine if the numerically and analytically modelled processes which cause evolution of the loop's transverse structure can be detected in observations.

As discussed in previous studies there are several shortcomings of the method used to infer the TDP of a coronal structure from the observed intensity profile. Since an isothermal approximation is made, i.e the temperature inside and outside the loop is assumed to be equal, temperature variation is detected as variation of the density, due to a variation of the instrumental response function. Any of the plasma which emits at temperatures not covered by the chosen AIA wavelength is not detected. The numerical data used  corresponds to a loop which is far from isothermal, however in Section 3 reasonable estimates for the density profile are obtained from the intensity profile in the AIA filter which corresponds to the core of the loop. Due to isothermal approximation in our method the radius ($R_L$) and inhomogeneous layer width ($\epsilon$) are underestimated, by $\approx$ 10 and 20\% respectively.

In Sect. \ref{sec:m1} the effect that downsampling the resolution and adding noise has on the final intensity profile of the loop was highlighted. A resolution of approximately 20 points across the loop, corresponding to a radius of $R_{L}$ = 3.5 Mm at AIA resolution, is seen to mask the appearance of the intensity peaks from the TWIKH rolls. This corresponds to the wider loops observed with AIA, and is therefore a best case scenario for current observations. Evolution of the loop in the TD maps can clearly be seen, as well as evolution of the density profile parameters inferred from the intensity profile. The main observational signatures when using Model $L$ for the density profile are; decreasing density enhancement ($A$), a widening inhomogeneous layer ($\epsilon$), a constant minor radius ($R_L$) and almost no visible transverse structuring. The visible decrease in intensity (and in the inferred value of $A$) was also detected in \cite{2017ApJ...836..219A}, and is due to the mixing of the internal and external plasma. The widening of the inhomogeneous layer was detected in \cite{2017ApJ...836..219A} as an increase in the loop's minor radius. Strong oscillatory behaviour is also seen in the time series of the TDP parameters, due to the effect of the oscillation itself on the TDP. 

In Sect. \ref{sec:m2m3} the effect of varying the width of the inhomogeneous layer and increasing the oscillation amplitude was investigated. The main difference in the former case was the delayed onset of the KHI despite the increased efficiency of the resonant absorption, as it takes longer for the sharp gradients in density and velocity to be generated. The variation of the TDP parameters after this onset is stronger however. The larger amplitude quickened the onset of the KHI and caused strong variation in the TDP of the loop, causing the appearance of multiple strands, visible even at the lower resolution $R2$. Both TD maps and time series for M2 and M3 are limited in length due to the numerical stability, the observational signatures are expected to be even stronger in reality. In \cite{2018ApJ...860...31P} the evolution of the inferred TDP over time for the analysed loop is presented, finding that the parameters showed some oscillatory behaviour, but no strong overall trend. However, this lack of KHI signatures could be due to the low oscillation amplitude. Further examples should be chosen and analysed in the same manner. The study and technique should also be extended to incorporate other EUV wavelengths or data from other instruments. This method for inferring the TDP is limited by the density profile used. The strong peaks in intensity generated in M3 meant that the uncertainties on the inferred parameters became large, as they can not be modelled by Model $L$.

For observational searches of KHI in oscillating loops it will be difficult to observe the TWIKW rolls directly. This is in part due to the unknown level of substructure within coronal loops. It is often difficult to determine if there are many threads within a given coronal loop, or if they are spatially separated along the line of sight \citep[e.g][]{2013A&A...556A.104P, 2016ApJ...826L..18B, 2017ApJ...840....4A}. The results presented here can be compared to observations of coronal loops which appear homogeneous in a given EUV channel, and it should be noted that a simple cool and dense loop model has been used. More detailed analysis of the observable signatures of the KHI in different observations was given in \cite{2017ApJ...836..219A}. Further work should also be done in the context of the numerical simulations performed. The timeseries for M2 and M3 could be extended with additional numerical treatment. Additionally, the effects of reconnection could be explored, due to the turbulence induced by the TWIKH rolls, and the effect of this on the observational signatures.

The changing width of the inhomogeneous layer highlighted has implications for the damping of the kink mode, as well as the seismology which is based on the damping behaviour. It also has implications for the spatial distribution of the energy disposition. Detection of significant evolution of coronal loop parameters during oscillations would increase the need for the inclusion of non-linear effects in observational analysis and theoretical modelling.

\section{Summary}

The development of KHI during kink oscillations causes a widening of the loop's inhomogeneous layer when approximated by a linear transition between the exterior and interior density. This may be detected observationally by inferring the density model parameters from the observed intensity profile using Bayesian inference. The main observational signatures  for an EUV channel corresponding to the loop core (for the loop model considered here) are ; decreasing intensity, a widening inhomogeneous layer, an unchanged radius, and visible transverse structuring, depending on how well resolved the loop is spatially. These effects occur slower for loops with wider inhomogeneous layers, and quicker for loops oscillating at higher amplitudes. These later two cases should also result in stronger observational signatures after the onset of the KHI, and visible transverse structuring appearing as multiple strands. 

A method to infer the transverse density profile (TDP) of coronal structures from the observed transverse intensity profile has been tested on numerical data. Despite the isothermal approximation made in the method, when applied to the non-isothermal loop in the numerical data the minor radius and inhomogeneous layer width were found to be underestimate by only $\approx$ 10 and 20\% respectively.

Future studies should search for the evolution of the transverse density profile of coronal loops in observational data. The potential effects of an evolving transverse density profile should also be considered in numerical and analytical studies as well as in seismology. The method for inferring the transverse density profile of the loops should be extended to include a transverse temperature structure, allowing observational studies to be extended to multiple EUV channels.

\acknowledgements
This work was supported by the European Research Council (ERC) under the SeismoSun Research Project No. 321141 (CRG, DJP) and by the British Council via the Institutional Links Programme (Project 277352569 - Seismology of Solar Coronal Active Regions) (CRG). DJP has received funding from the ERC under the European Union's Horizon 2020 research and innovation programme (grant agreement No 724326).  P.A. has received funding from the UK Science and Technology Facilities Council (Consolidated Grant ST/K000950/1), the European Union Horizon 2020 research and innovation programme (grant agreement No. 647214) and his STFC Ernest Rutherford Fellowship (grant agreement No. ST/R004285/1). Numerical computations were carried out on Cray XC30 at the Center for Computational Astrophysics, NAOJ. We thank S. Anfinogentov for providing the MCMC Bayesian inference code, and I. De Moortel for useful comments and discussion.

\bibliography{references.bib}

\begin{thebibliography}{}
\expandafter\ifx\csname natexlab\endcsname\relax\def\natexlab#1{#1}\fi
\providecommand{\url}[1]{\href{#1}{#1}}

\bibitem[{{Antolin} {et~al.}(2016){Antolin}, {De Moortel}, {Van Doorsselaere},
  \& {Yokoyama}}]{2016ApJ...830L..22A}
{Antolin}, P., {De Moortel}, I., {Van Doorsselaere}, T., \& {Yokoyama}, T.
  2016, \apjl, 830, L22

\bibitem[{{Antolin} {et~al.}(2017){Antolin}, {De Moortel}, {Van Doorsselaere},
  \& {Yokoyama}}]{2017ApJ...836..219A}
---. 2017, \apj, 836, 219

\bibitem[{{Antolin} {et~al.}(2015){Antolin}, {Okamoto}, {De Pontieu},
  {Uitenbroek}, {Van Doorsselaere}, \& {Yokoyama}}]{2015ApJ...809...72A}
{Antolin}, P., {Okamoto}, T.~J., {De Pontieu}, B., {et~al.} 2015, \apj, 809, 72

\bibitem[{{Antolin} {et~al.}(2014){Antolin}, {Yokoyama}, \& {Van
  Doorsselaere}}]{2014ApJ...787L..22A}
{Antolin}, P., {Yokoyama}, T., \& {Van Doorsselaere}, T. 2014, \apjl, 787, L22

\bibitem[{{Arregui}(2017)}]{2017arXiv170908372A}
{Arregui}, I. 2017, ArXiv e-prints, arXiv:1709.08372

\bibitem[{{Arregui} \& {Asensio Ramos}(2011)}]{2011ApJ...740...44A}
{Arregui}, I., \& {Asensio Ramos}, A. 2011, \apj, 740, 44

\bibitem[{{Arregui} {et~al.}(2015){Arregui}, {Soler}, \& {Asensio
  Ramos}}]{2015ApJ...811..104A}
{Arregui}, I., {Soler}, R., \& {Asensio Ramos}, A. 2015, \apj, 811, 104

\bibitem[{{Aschwanden} \& {Boerner}(2011)}]{2011ApJ...732...81A}
{Aschwanden}, M.~J., \& {Boerner}, P. 2011, \apj, 732, 81

\bibitem[{{Aschwanden} {et~al.}(1999){Aschwanden}, {Fletcher}, {Schrijver}, \&
  {Alexander}}]{1999ApJ...520..880A}
{Aschwanden}, M.~J., {Fletcher}, L., {Schrijver}, C.~J., \& {Alexander}, D.
  1999, \apj, 520, 880

\bibitem[{{Aschwanden} \& {Peter}(2017)}]{2017ApJ...840....4A}
{Aschwanden}, M.~J., \& {Peter}, H. 2017, \apj, 840, 4

\bibitem[{{Brooks} {et~al.}(2016){Brooks}, {Reep}, \&
  {Warren}}]{2016ApJ...826L..18B}
{Brooks}, D.~H., {Reep}, J.~W., \& {Warren}, H.~P. 2016, \apjl, 826, L18

\bibitem[{{Browning} \& {Priest}(1984)}]{1984A&A...131..283B}
{Browning}, P.~K., \& {Priest}, E.~R. 1984, \aap, 131, 283

\bibitem[{{Clack} \& {Ballai}(2009)}]{2009PhPl...16g2115C}
{Clack}, C.~T.~M., \& {Ballai}, I. 2009, Physics of Plasmas, 16, 072115

\bibitem[{{De Moortel} \& {Nakariakov}(2012)}]{2012RSPTA.370.3193D}
{De Moortel}, I., \& {Nakariakov}, V.~M. 2012, Royal Society of London
  Philosophical Transactions Series A, 370, 3193

\bibitem[{{De Moortel} {et~al.}(2016){De Moortel}, {Pascoe}, {Wright}, \&
  {Hood}}]{0741-3335-58-1-014001}
{De Moortel}, I., {Pascoe}, D.~J., {Wright}, A.~N., \& {Hood}, A.~W. 2016,
  Plasma Physics and Controlled Fusion, 58, 014001.
\newblock \url{http://stacks.iop.org/0741-3335/58/i=1/a=014001}

\bibitem[{{Goddard} \& {Nakariakov}(2016)}]{2016A&A...590L...5G}
{Goddard}, C.~R., \& {Nakariakov}, V.~M. 2016, \aap, 590, L5

\bibitem[{{Goddard} {et~al.}(2016){Goddard}, {Nistic{\`o}}, {Nakariakov}, \&
  {Zimovets}}]{2016A&A...585A.137G}
{Goddard}, C.~R., {Nistic{\`o}}, G., {Nakariakov}, V.~M., \& {Zimovets}, I.~V.
  2016, \aap, 585, A137

\bibitem[{{Goddard} {et~al.}(2017){Goddard}, {Pascoe}, {Anfinogentov}, \&
  {Nakariakov}}]{2017A&A...605A..65G}
{Goddard}, C.~R., {Pascoe}, D.~J., {Anfinogentov}, S., \& {Nakariakov}, V.~M.
  2017, \aap, 605, A65

\bibitem[{{Grigis} {et~al.}(2013){Grigis}, {Yingna}, \& {Weber}}]{aiapsf}
{Grigis}, P., {Yingna}, S., \& {Weber}, M. 2013, Tech. Rep., AIA team.
\newblock
  \url{http://hesperia.gsfc.nasa.gov/ssw/sdo/aia/idl/psf/DOC/psfreport.pdf}

\bibitem[{{Handy} {et~al.}(1999){Handy}, {Acton}, {Kankelborg}, {Wolfson},
  {Akin}, {Bruner}, {Caravalho}, {Catura}, {Chevalier}, {Duncan}, {Edwards},
  {Feinstein}, {Freeland}, {Friedlaender}, {Hoffmann}, {Hurlburt}, {Jurcevich},
  {Katz}, {Kelly}, {Lemen}, {Levay}, {Lindgren}, {Mathur}, {Meyer}, {Morrison},
  {Morrison}, {Nightingale}, {Pope}, {Rehse}, {Schrijver}, {Shine}, {Shing},
  {Strong}, {Tarbell}, {Title}, {Torgerson}, {Golub}, {Bookbinder}, {Caldwell},
  {Cheimets}, {Davis}, {Deluca}, {McMullen}, {Warren}, {Amato}, {Fisher},
  {Maldonado}, \& {Parkinson}}]{1999SoPh..187..229H}
{Handy}, B.~N., {Acton}, L.~W., {Kankelborg}, C.~C., {et~al.} 1999, \solphys,
  187, 229

\bibitem[{{Hood} {et~al.}(2013){Hood}, {Ruderman}, {Pascoe}, {De Moortel},
  {Terradas}, \& {Wright}}]{2013A&A...551A..39H}
{Hood}, A.~W., {Ruderman}, M., {Pascoe}, D.~J., {et~al.} 2013, \aap, 551, A39

\bibitem[{{Howson} {et~al.}(2017){Howson}, {De Moortel}, \&
  {Antolin}}]{2017A&A...602A..74H}
{Howson}, T.~A., {De Moortel}, I., \& {Antolin}, P. 2017, \aap, 602, A74

\bibitem[{{Kudoh} \& {Shibata}(1999)}]{1999ApJ...514..493K}
{Kudoh}, T., \& {Shibata}, K. 1999, \apj, 514, 493

\bibitem[{{Lemen} {et~al.}(2012){Lemen}, {Title}, {Akin}, {Boerner}, {Chou},
  {Drake}, {Duncan}, {Edwards}, {Friedlaender}, {Heyman}, {Hurlburt}, {Katz},
  {Kushner}, {Levay}, {Lindgren}, {Mathur}, {McFeaters}, {Mitchell}, {Rehse},
  {Schrijver}, {Springer}, {Stern}, {Tarbell}, {Wuelser}, {Wolfson}, {Yanari},
  {Bookbinder}, {Cheimets}, {Caldwell}, {Deluca}, {Gates}, {Golub}, {Park},
  {Podgorski}, {Bush}, {Scherrer}, {Gummin}, {Smith}, {Auker}, {Jerram},
  {Pool}, {Soufli}, {Windt}, {Beardsley}, {Clapp}, {Lang}, \&
  {Waltham}}]{2012SoPh..275...17L}
{Lemen}, J.~R., {Title}, A.~M., {Akin}, D.~J., {et~al.} 2012, \solphys, 275, 17

\bibitem[{{Li} {et~al.}(2017){Li}, {Liu}, \& {Vai Tam}}]{2017ApJ...842...99L}
{Li}, H., {Liu}, Y., \& {Vai Tam}, K. 2017, \apj, 842, 99

\bibitem[{{Long} {et~al.}(2017){Long}, {Valori}, {P{\'e}rez-Su{\'a}rez},
  {Morton}, \& {V{\'a}squez}}]{2017A&A...603A.101L}
{Long}, D.~M., {Valori}, G., {P{\'e}rez-Su{\'a}rez}, D., {Morton}, R.~J., \&
  {V{\'a}squez}, A.~M. 2017, \aap, 603, A101

\bibitem[{{Magyar} \& {Van
  Doorsselaere}(2016{\natexlab{a}})}]{2016A&A...595A..81M}
{Magyar}, N., \& {Van Doorsselaere}, T. 2016{\natexlab{a}}, \aap, 595, A81

\bibitem[{{Magyar} \& {Van
  Doorsselaere}(2016{\natexlab{b}})}]{2016ApJ...823...82M}
---. 2016{\natexlab{b}}, \apj, 823, 82

\bibitem[{{Montes-Sol{\'{\i}}s} \& {Arregui}(2017)}]{2017ApJ...846...89M}
{Montes-Sol{\'{\i}}s}, M., \& {Arregui}, I. 2017, \apj, 846, 89

\bibitem[{{Nakariakov} \& {Ofman}(2001)}]{2001A&A...372L..53N}
{Nakariakov}, V.~M., \& {Ofman}, L. 2001, \aap, 372, L53

\bibitem[{{Nakariakov} {et~al.}(1999){Nakariakov}, {Ofman}, {Deluca},
  {Roberts}, \& {Davila}}]{1999Sci...285..862N}
{Nakariakov}, V.~M., {Ofman}, L., {Deluca}, E.~E., {Roberts}, B., \& {Davila},
  J.~M. 1999, Science, 285, 862

\bibitem[{{Ofman} {et~al.}(1994){Ofman}, {Davila}, \&
  {Steinolfson}}]{1994GeoRL..21.2259O}
{Ofman}, L., {Davila}, J.~M., \& {Steinolfson}, R.~S. 1994, \grl, 21, 2259

\bibitem[{{Pagano} \& {De Moortel}(2017)}]{2017A&A...601A.107P}
{Pagano}, P., \& {De Moortel}, I. 2017, \aap, 601, A107

\bibitem[{{Pascoe} {et~al.}(2017{\natexlab{a}}){Pascoe}, {Anfinogentov},
  {Nistic{\`o}}, {Goddard}, \& {Nakariakov}}]{2017A&A...600A..78P}
{Pascoe}, D.~J., {Anfinogentov}, S., {Nistic{\`o}}, G., {Goddard}, C.~R., \&
  {Nakariakov}, V.~M. 2017{\natexlab{a}}, \aap, 600, A78

\bibitem[{{Pascoe} {et~al.}(2018){Pascoe}, {Anfinogentov}, {Goddard}, \&
  {Nakariakov}}]{2018ApJ...860...31P}
{Pascoe}, D.~J., {Anfinogentov}, S.~A., {Goddard}, C.~R., \& {Nakariakov},
  V.~M. 2018, \apj, 860, 31

\bibitem[{{Pascoe} {et~al.}(2017{\natexlab{b}}){Pascoe}, {Goddard},
  {Anfinogentov}, \& {Nakariakov}}]{2017A&A...600L...7P}
{Pascoe}, D.~J., {Goddard}, C.~R., {Anfinogentov}, S., \& {Nakariakov}, V.~M.
  2017{\natexlab{b}}, \aap, 600, L7

\bibitem[{{Pascoe} {et~al.}(2016){Pascoe}, {Goddard}, {Nistic{\`o}},
  {Anfinogentov}, \& {Nakariakov}}]{2016A&A...589A.136P}
{Pascoe}, D.~J., {Goddard}, C.~R., {Nistic{\`o}}, G., {Anfinogentov}, S., \&
  {Nakariakov}, V.~M. 2016, \aap, 589, A136

\bibitem[{{Pascoe} {et~al.}(2013){Pascoe}, {Hood}, {De Moortel}, \&
  {Wright}}]{2013A&A...551A..40P}
{Pascoe}, D.~J., {Hood}, A.~W., {De Moortel}, I., \& {Wright}, A.~N. 2013,
  \aap, 551, A40

\bibitem[{{Peter} {et~al.}(2013){Peter}, {Bingert}, {Klimchuk}, {de Forest},
  {Cirtain}, {Golub}, {Winebarger}, {Kobayashi}, \&
  {Korreck}}]{2013A&A...556A.104P}
{Peter}, H., {Bingert}, S., {Klimchuk}, J.~A., {et~al.} 2013, \aap, 556, A104

\bibitem[{{Ruderman} {et~al.}(2017){Ruderman}, {Shukhobodskiy}, \&
  {Erd{\'e}lyi}}]{2017A&A...602A..50R}
{Ruderman}, M.~S., {Shukhobodskiy}, A.~A., \& {Erd{\'e}lyi}, R. 2017, \aap,
  602, A50

\bibitem[{{Sarkar} {et~al.}(2016){Sarkar}, {Pant}, {Srivastava}, \&
  {Banerjee}}]{2016SoPh..291.3269S}
{Sarkar}, S., {Pant}, V., {Srivastava}, A.~K., \& {Banerjee}, D. 2016,
  \solphys, 291, 3269

\bibitem[{{Soler} {et~al.}(2010){Soler}, {Terradas}, {Oliver}, {Ballester}, \&
  {Goossens}}]{2010ApJ...712..875S}
{Soler}, R., {Terradas}, J., {Oliver}, R., {Ballester}, J.~L., \& {Goossens},
  M. 2010, \apj, 712, 875

\bibitem[{{Terradas} {et~al.}(2008){Terradas}, {Andries}, {Goossens},
  {Arregui}, {Oliver}, \& {Ballester}}]{2008ApJ...687L.115T}
{Terradas}, J., {Andries}, J., {Goossens}, M., {et~al.} 2008, \apjl, 687, L115

\bibitem[{{Terradas} {et~al.}(2017){Terradas}, {Magyar}, \& {Van
  Doorsselaere}}]{2017arXiv171206955T}
{Terradas}, J., {Magyar}, N., \& {Van Doorsselaere}, T. 2017, ArXiv e-prints,
  arXiv:1712.06955

\bibitem[{{Terradas} \& {Ofman}(2004)}]{2004ApJ...610..523T}
{Terradas}, J., \& {Ofman}, L. 2004, \apj, 610, 523

\bibitem[{{Uchimoto} {et~al.}(1991){Uchimoto}, {Strauss}, \&
  {Lawson}}]{1991SoPh..134..111U}
{Uchimoto}, E., {Strauss}, H.~R., \& {Lawson}, W.~S. 1991, \solphys, 134, 111

\bibitem[{{Van Doorsselaere} {et~al.}(2004){Van Doorsselaere}, {Andries},
  {Poedts}, \& {Goossens}}]{2004ApJ...606.1223V}
{Van Doorsselaere}, T., {Andries}, J., {Poedts}, S., \& {Goossens}, M. 2004,
  \apj, 606, 1223

\bibitem[{{Vasheghani Farahani} {et~al.}(2012){Vasheghani Farahani},
  {Nakariakov}, {Verwichte}, \& {Van Doorsselaere}}]{2012A&A...544A.127V}
{Vasheghani Farahani}, S., {Nakariakov}, V.~M., {Verwichte}, E., \& {Van
  Doorsselaere}, T. 2012, \aap, 544, A127

\bibitem[{{Verwichte} {et~al.}(2013){Verwichte}, {Van Doorsselaere}, {White},
  \& {Antolin}}]{2013A&A...552A.138V}
{Verwichte}, E., {Van Doorsselaere}, T., {White}, R.~S., \& {Antolin}, P. 2013,
  \aap, 552, A138

\bibitem[{{White} \& {Verwichte}(2012)}]{2012A&A...537A..49W}
{White}, R.~S., \& {Verwichte}, E. 2012, \aap, 537, A49

\end{thebibliography}

\end{document}